\documentclass[review,3p,10pt]{elsarticle}
\usepackage{bm}
\usepackage{lineno,xcolor}
\usepackage{enumitem}
\usepackage{textcomp}
\usepackage{appendix}

\begin{document}

\begin{frontmatter}
\title{Polarization dependence in inelastic scattering of electrons
by hydrogen atoms in a circularly polarized laser field}
\author{Gabriela Buic\u{a}}
\ead{buica@spacescience.ro}
\address{Institute of Space Science, P.O. Box MG-23, Ro 77125,
Bucharest-M\u{a}gurele, Romania}

\begin{abstract}
 We theoretically study the influence of laser polarization in
inelastic scattering of  electrons by hydrogen atoms
 in the presence of a circularly polarized laser field
 in the domain of field strengths below $10^{7}$ V/cm and high projectile energies.
A semi-perturbative approach is used in which the interaction of
 the projectile electrons  with the laser field is described by
Gordon-Volkov wave functions,  while the interaction of the hydrogen atom
 with the laser field is described by first-order  time-dependent perturbation theory.
A \textit{closed analytical solution} is derived in laser-assisted
 inelastic electron-hydrogen scattering  for the $1s \to nl$
excitation cross section  which is valid for both circular and linear polarizations.
 For the  excitation of the $n=2$ levels  simple analytical expressions of
differential cross section  are derived for laser-assisted inelastic scattering
 in the perturbative domain, and the   differential cross sections by the
 circularly and linearly polarized  laser fields and their ratios for
one- and two-photon absorption  are calculated as a function of the scattering angle.
Detailed numerical results  for the angular dependence and the
resonance structure of  the differential cross sections  is discussed
 for the $1s \to 4l$ excitations of  hydrogen in a  circularly polarized laser field.
\end{abstract}
\date{\today}

\end{frontmatter}

\section{Introduction}
\label{I}

It is well known that the atomic processes which take place in the presence of an  electromagnetic field and
involve atomic spectra exhibit field polarization-dependent characteristics \cite{cohen}.
Obviously polarization dependence effects occur as well for electron-atom scattering in the
presence of a laser field, which is of potential interest in applied domains
such as laser and plasma physics \cite{plasma},  astrophysics \cite{astro},
 or fundamental atomic collision theory \cite{massey}.
Detailed reports on the laser-assisted electron-atom
collisions can be found in several review papers
 \cite{mason,ehl1998,ehl2001} and books \cite{bransden,joa2012}.
The polarization dependence of elastic laser-assisted electron-hydrogen scattering
was first theoretically studied for circularly polarized (CP) fields and  differences between the
angular distributions for linear and circular polarizations were reported
by Fainstein and Maquet \cite{fainstein},
by considering a model potential for the calculation of the atomic wave functions.
Theoretical studies involving CP fields with  the atomic dressing taken
into account in second-order of time-dependent perturbation theory (TDPT) were performed
for elastic case by Cionga and coworkers \cite{acgabi1,acgabi2}, and
the dichroism  of the scattered electrons was investigated
in elastic electron-hydrogen scattering by circularly and elliptically polarized
laser fields  \cite{acgabiopt}.
The  polarization effects  in laser-assisted  electron-atom inelastic collisions
for the  excitation of the $n=2$ and $3$ states in H and He,
 were reported \cite{akramine99}
for two specific scattering geometries
 where the wave vector of the CP field is coplanar and parallel to the scattering plane.
A different method based on a discrete basis of Sturmian functions was employed  to calculate the relevant
 atomic terms of the scattering matrix and, in contrast to this previous treatment,
a major purpose of our  paper is to provide  new analytical formulas needed to  compute the
 differential cross section (DCS) for the  laser-assisted excitation collisions $e^{-} + H({1s})$ by CP laser fields,
  that allow further investigations of the polarization effect and give more insight into  the scattering process.
It is well known that by increasing the atomic excitation the
laser-dressing effects induced by the dipole polarizability \cite{Radzig}  should
increase in importance \cite{milo,acgabi3,gabi2s}, and therefore with an increasing
probability of experimental detection.
As far as we known the first theoretical works on the laser-assisted inelastic scattering of electrons by
 atoms  taking into account the atomic ``dressing''
 (i.e., the dipole distortion of the atom by the laser field) were performed
in different approaches such as first-order perturbation theory \cite{Jetzke,Francken,Bhattacharya,ac},
 non-perturbative Floquet theory  \cite{vucic95},
 or relativistic method \cite{voitkiv}.

Similarly to the case of elastic collisions, most experimental investigations of
inelastic electron-atom scattering are focused on  collision processes in linearly
polarized (LP) laser fields \cite{mason}.
 The first experimental observation of an inelastic process
involving one-photon absorption was made by Mason and Newell for electron-helium scattering
in elliptically \cite{mason89} and  circularly \cite{mason90} polarized laser fields.
In the recent years due to progresses in new experimental techniques there is  a renewed interest
 in investigating  electron-atom scattering in the presence of a laser field  \cite{musa,Kanya,Harak,Kanya2},
and therefore there is a need for simple analytical formulas in order to describe the process.
Very recently, there is also an increased interest in the  simultaneous electron-photon excitation
 of helium in the presence of a laser field   using a nonperturbative \textit{R}-matrix Floquet theory
  \cite{Dunseath2011}, or involving a semi-perturbative method in the second-order Born approximation \cite{Ajana}.

In this paper we  investigate the $1s\rightarrow nl$  inelastic scattering
of fast electrons  by hydrogen atoms in their ground states in the presence of a CP laser field.
In Sec. \ref{II}  is described the semi-perturbative framework used
to  derive the analytical formulas for the DCS for excitation of an arbitrary state.
Because the scattering process under investigation is a very complex problem, the
theoretical approach poses considerable difficulties and several assumptions are made.
(i) Moderate laser field strengths below $10^{7}$ V/cm and
 fast projectile electrons are considered in order
to safely  neglect the second-order Born approximation
in the scattering potential as well as the exchange scattering \cite{b-j,Francken,Makhoute}.
(ii) The interaction of  the projectile electrons with the laser field
is described by a Gordon-Volkov wave function \cite{volkov}.
(iii) The dressing of the hydrogen atom by the laser field  is described
within the first-order time-dependent perturbation theory (TDPT) in the
field  \cite{vf1}.
This theoretical approach is similar to the one that has been  used  to a related problem
 of laser-assisted excitation collisions $e^{-} + H({2s})$ by LP laser fields \cite{gabi2s}
 and,  contrasting with this  work, the main differences reside in the initial atomic state, that is now ${1s}$,
 and the circular polarization of the laser field, which result in different specific expressions
 for the  projectile and atomic contributions to DCS by CP fields.
In comparison to the earlier theoretical works  \cite{Francken,ac,akramine99}
the present semi-perturbative approach provides, for the first-time as far as we know,
 a closed analytical form for the DCS, which
includes the atomic dressing effects, and  is valid for both LP and CP fields.
We obtain  \textit{useful analytic formulas} for DCSs
in the laser-assisted inelastic scattering  in the perturbative  limit
 as well  in the soft-photon limit for the $1s \to nl$ excitation.
 \textit{Simple asymptotic formulas} for DCS are derived at low-photon energies
for the $1s \to 2s$  and $1s \to 2p$ excitations.
Section \ref{III} is devoted to discussion of the numerical results,  where
 the angular distributions of the DCSs by CP fields are presented for
 the excitation of $n=2$ and $4$ subshells,  and the resonance structure of the
DCSs as a function of the photon energy is analyzed for the  excitation of $n=4$ subshells.
The  ratios of the differential cross sections by the
 CP and LP laser fields are calculated as a function of the scattering angle
 for  the excitation of $n=2$ subshells  accompanied by  one- and two-photon absorption,
 which could be very useful from the experimental point of view.
Atomic units (a.u.) are used throughout this paper unless otherwise specified.

\section{Semi-perturbative scattering theory at moderate laser intensities and circular polarizations}
\label{II}

We consider the following laser-assisted inelastic scattering of electrons by  hydrogen atoms in their ground state
(free-free transitions):

\begin{equation}
e^-(E_{k_i},{\mathbf{k}_i})+ {\rm H}(1s)
+ N_i \; \gamma \;(\omega, \boldsymbol{\varepsilon})
\to
e^-(E_{k_f},{\mathbf{k}_f})+ {\rm H}(nlm)
+ N_f \; \gamma \;(\omega, \boldsymbol{\varepsilon}),
\label{process}
\end{equation}

\noindent
where $ E_{k_i} \;(E_{k_f}) $ and $\mathbf{k}_{i} \;(\mathbf{k}_{f}$) denote
the kinetic energy and  the momentum vector of the projectile electron in its initial
(final) state.
 $\gamma$ denotes a photon with the energy $\omega $ and the  polarization
vector $\boldsymbol{\varepsilon}$.
The hydrogen atom is initially in its ground state  and finally is excited to a final
state defined by the quantum numbers $n, l$, and $m$.
The above scattering process is considered {\it inelastic} since
the initial and final states of the target are not identical and therefore
the projectile electron energies satisfy the conservation relation:
$ E_{k_f}   =   E_{k_i} + {E}_{1s} - { E}_{n} + N \omega$, where
 ${ E}_{1s}$ and ${E}_{n}$ are the energies of the  ground and  excited states,
and  $N =N_i-N_f$  denotes the net number of exchanged photons
between the colliding system and the laser field.
The kinetic energy spectrum of the scattered electrons consists of a
\textit{central line} that corresponds to $N=0$
and a number of  symmetrically located \textit{sideband} lines with
$|N|=  1, 2, 3,$..., equally spaced by the photon energy \cite{mason}.
The laser field  is assumed to be a monochromatic electric field and is treated classically as
\begin{equation}
 \mbox{\boldmath $\cal E$} (t) =
 \frac{i}{2} {\cal E}_0 \boldsymbol{\varepsilon}  \exp(-i  \omega  t) + cc,
\label{field}
\end{equation}

\noindent
where ${\cal E}_0$ is the peak amplitude of the electric field, and
the polarization vector is given by
\begin{equation}
 \boldsymbol{\varepsilon} = \cos( \xi/2)[ \mathbf{e}_{i}+  i \mathbf{e}_{j} \tan ( \xi/2)],
\end{equation}
\noindent
where $\mathbf{e}_{i}$ and  $\mathbf{e}_{j}$ are the unit vectors along two orthogonal directions
 in the polarization plane and $\xi$ denotes the ellipticity degree.
In particular, the ellipticity  is
$\xi =\pm \pi/2$ for a  CP field and
 $\xi =0$ for a LP field.

\subsection{Scattering matrix in the semi-perturbative domain}

As stated previously, in the present work we consider fast projectile electrons
 such that the scattering  process can be well treated within the first-order
Born approximation in the scattering potential
\begin{equation}
V(r,R)=-\frac 1R+\frac 1{|\bf{R}-\bf{r}|},
\label{pot}
\end{equation}
\noindent
where $\mathbf{R}$ and $\mathbf{r}$  denote the position vectors of the projectile and bound electrons,
and we use a semi-perturbative approach for the
scattering process similar to the one developed by Byron and Joachain \cite{b-j}.
We start with the calculation of  the scattering matrix element \cite{Francken} in the domain of
  high scattering energies as
\begin{equation}
S_{fi} = -i \int_{-\infty}^{\infty} dt
\langle \chi_{ \mathbf{k}_f}( \mathbf{R},t)
\Psi_{n_fl_fm_f}(\mathbf{r},t)
 		 |V( \mathbf{r},\mathbf{R}) |
 	      {\chi}_{ \mathbf{k}_i}( \mathbf{R}, t)
\Psi_{n_il_im_i}(\mathbf{r},t)
			\rangle
\;,
\label{sm}
\end{equation}
\noindent
where $\chi_{\mathbf{k}_i}$  and $\chi_{\mathbf{k}_f}$ represent  the initial and final
Gordon-Volkov wave functions  of the projectile electron in the  laser field,
 and $\Psi_{n_il_im_i} $  and $\Psi_{n_fl_fm_f}$ are the
initial and  final wave functions of the bound electron in the  laser field,
which are given by Eqs. (3) and (5) of Ref. \cite{gabi2s}.
For fast projectile electrons ($E_{k_i}>150$ eV) the exchange effects are safely neglected
\cite{Francken} and are not included in the calculation of the scattering matrix.
As long as the laser field strength remains moderate the interaction of
the hydrogen atom  with the laser field is considered within the first-order TDPT.
More details about the dressing effects are discussed  elsewhere \cite{acgabi1,acgabi2,acgabi3}.
By using the  Jacob-Anger expansion of the exponential
\begin{equation}
\exp{[i X  \sin(\omega  t)]} =
\sum_{N=-\infty}^{+\infty} J_N(X) \exp{(i N \omega t)},
\end{equation}

\noindent
we develop  the Gordon-Volkov wave functions
in terms of the ordinary Bessel functions  \cite{Watson} of order $N$, $J_N$,
\noindent
\begin{equation}
\exp{[-i  \bm{\alpha}(t)   \cdot \mathbf{k} ]} =
\sum_{N=-\infty}^{+\infty} J_N(X_k) \exp{(-i N \omega t +i N \phi_k)},
\label{bgv}
\end{equation}
\noindent
where $ X_k=\alpha_0| \boldsymbol{\varepsilon}\cdot \mathbf{k} |$
 is the argument of the Bessel function
and the dynamical phase $\phi_k$ is defined through the expression
$\exp{(i \phi_k)} = | \boldsymbol{\varepsilon}\cdot \mathbf{k}| / (\boldsymbol{\varepsilon}\cdot \mathbf{k}) $.
The quiver motion of the projectile electron in the electric field
  $\mbox{\boldmath $\cal E$} (t)$, Eq. (\ref{field}), is described by
\begin{equation}
\boldsymbol{\alpha} (t) ={\alpha_0}
[ \mathbf{e}_{i} \sin (\omega t) \cos (\xi/2) - \mathbf{e}_{i} \cos (\omega t) \sin (\xi/2)] ,
\label{quiver}
\end{equation}
 with the amplitude   ${\alpha_0} = {\cal E}_0/ \omega ^{2}$.
For a CP field
$ X_k^{CP}=(\alpha_0/\sqrt{2}) \sqrt{(\mathbf{e}_i\cdot \mathbf{k})^2+(\mathbf{e}_j\cdot \mathbf{k})^2}$
and $\phi_k^{CP}=\arctan{(\mathbf{e}_j\cdot \mathbf{k})/(\mathbf{e}_i\cdot \mathbf{k})} +s\pi$,
while for a LP field
$ X_k^{LP}=\alpha_0 | \mathbf{e}_i\cdot \mathbf{k} |$ and $\phi_k^{LP}=s\pi$, where $s$ is an integer.
\noindent
By substituting the projectile and atomic wave functions into
 Eq. (\ref{sm}) we obtain, after integrating over time and
 projectile coordinate, the scattering matrix
 $ S_{fi}$ for the laser-assisted  inelastic electron-hydrogen collisions
accompanied by the exchange of $N$ photons (absorbed or emitted),

\begin{equation}
S_{fi} (N)  =
- 2\pi i \sum_{N=-\infty}^{+\infty}	T_{nlm}(N) \;
\delta( E_{k_f} + { E}_{n} -   E_{k_i} - { E}_{1s}  - N \omega) \;,
\label{smt}
\end{equation}

\noindent
where $\delta$ is the Dirac delta function which is related to the  energy conservation
$ E_{k_f}=  E_{k_i} +  { E}_{1s} -{ E}_{n}  + N \omega $.
 $ T_{nlm} (N) $ denotes the total transition amplitude
for the  scattering  process (1), which  can be split  as a sum of two terms:

\begin{equation}
T_{nlm}(N) = T_{nlm}^{(0)}(N) + T_{nlm}^{(1)}(N)
.\label{tgen}
\end{equation}
\noindent
The first term on the right-hand side of Eq. (\ref{tgen}),  $T_{nlm}^{(0)}$,
represents the inelastic  transition amplitude due to projectile electron,
and  describes the direct excitation of hydrogen  by projectile electron interaction,

\begin{equation}
T_{nlm}^{(0)}(N) =
J_N ( X_{q} )
\langle\psi_{nlm}|F(\mathbf{q})|\psi_{100}\rangle ,
\label{tn0}
\end{equation}

\noindent
where $ \psi _{nlm} $
is the unperturbed excited-state wave function of the bound electron,
$  X_{q}=\alpha_0| \boldsymbol{\varepsilon}\cdot {\mathbf{q}}|$, and
  $ \mathbf{q}= \mathbf{k}_i - \mathbf{k}_f$ denotes the momentum transfer vector
of the projectile electron.
 $F(\mathbf{q})$ represents the generalized form factor given by

\begin{equation}
F(\mathbf{q}) = \frac{1}{2 \pi^2 q^2}
	\left[ \exp{(i  \mathbf{q} \cdot \mathbf{r}) } - \delta_{n1}\delta_{l0}\delta_{m0}  \right].
\label{ff}
\end{equation}

\noindent
The laser and the projectile  contributions to the transition amplitude
$T_{nlm}^{(0)}$, are completely decoupled   in Eq. (\ref{tn0}) because the
laser-field dependence of the electronic transition amplitude
  is included in  the argument of the Bessel function only.
We recall that the Bunkin-Fedorov \cite{bf} and  Kroll-Watson \cite{kw}   formulas are
inappropriate for laser-assisted inelastic electron-atom scattering
(in which the atom changes its initial state)
since both   approximations are obtained for  a static potential.

\noindent
The second term on the right-hand side of the inelastic transition amplitude
 Eq. (\ref{tgen}), $T_{nlm}^{(1)}$, represents the atomic transition amplitude.
It is related to  the dressing of the atomic state by the laser field which is described by
 the first-order radiative correction  defined in Ref. \cite{vf1}, and its evaluation is
 more complicated than  $T_{nlm}^{(0)}$.
 Briefly, after some algebra the atomic transition amplitude $T_{nlm}^{(1)}$ is expressed by
\begin{equation}
T^{(1)}_{nlm}(N) =
-\frac {\alpha_0   \omega}{2}
\left[
J_{N-1}(  X_{q}  )
{\cal M}_{at}^{(1)}(\Omega_1^+,\Omega_n^-,\mathbf{q})\; e^{-i \phi_q}
 +
J_{N+1}(  X_{q}  )
{\cal M}_{at}^{(1)}(\Omega_1^-,\Omega_n^+, \mathbf{q})\; e^{i \phi_q}
\right]
,\label{tn1}
\end{equation}
where $ {\cal M}_{at}^{(1)}(\Omega_1^+,\Omega_n^-,\mathbf{q})$
and ${\cal M}_{at}^{(1)}(\Omega_1^-,\Omega_n^+,\mathbf{q})$
are specific first-order atomic transition amplitudes
\begin{equation}
{\cal M}_{at}^{(1)}\left( \Omega_1^+,\Omega_n^-,\mathbf{q}\right) =
\langle\psi _{nlm}|F(\mathbf{q})| \boldsymbol{\varepsilon}
\cdot
{\mathbf{w}}_{100}( \Omega_1^{+}) \rangle
+\langle  \boldsymbol{\varepsilon}^*\cdot {\mathbf{w}}_{nlm}
( \Omega_n^{-} )|F(\mathbf{q})|\psi _{100}\rangle,
\label{def1}
\end{equation}
that corresponds to one-photon absorption, and
\begin{equation}
{\cal M}_{at}^{(1)}\left(\Omega_1^-,\Omega_n^+,\mathbf{q}\right) =
\langle\psi _{nlm}|F(\mathbf{q})| \boldsymbol{\varepsilon}^*\cdot
{\mathbf{w}}_{100}( \Omega_1^{-}) \rangle
+\langle  \boldsymbol{\varepsilon}\cdot {\mathbf{w}}_{nlm}
( \Omega_n ^{+})
|F(\mathbf{q})|\psi_{100}\rangle,
\label{def2}
\end{equation}
that is related to one-photon emission.
  ${\bf{w}}_{nlm}$ are the  linear-response vectors  \cite{vf1}  which depend on the energies
$ \Omega_n^{\pm} = {E}_n \pm \omega$.
Obviously, only one of the $N$ photons is exchanged between the laser field and  the bound electron.
The first term on  the right-hand side of both Eqs. (\ref{def1}) and  (\ref{def2})
describes  first the atom interacting  with the laser field followed by
 the   projectile electron-atom interaction, while
in the second term the projectile electron-atom interaction precedes the laser-atom interaction.

\subsection{Differential cross section for excitation by a CP field}
\label{DCSe}

After performing the angular integration in Eq. (\ref{tn0})
the electronic transition amplitude $T_{nlm}^{(0)}$ takes the form
\begin{equation}
T_{nlm}^{(0)} (N) = \frac{1}{(2\pi)^2}
J_N(X_q) f_{el}^{B1}(q),
\label{tne}
\end{equation}
where $ f_{el}^{B1}(q)$ describes the first-order Born approximation of the
scattering amplitude for field-free inelastic process.
The evaluation of $ f_{el}^{B1}(q)$ for excitation of hydrogen atom from its ground state gives
\begin{equation}
 f_{el}^{B1}(q)= -
\frac{ 4  \sqrt{\pi} \; i^l}{q^{2} }
\; Y^*_{lm}(\hat{\mathbf{q}}) {\cal I}_{nl}(q) ,
\label{fn0i}
\end{equation}
where $ \hat{\mathbf{q}} = \mathbf{q}/|  \mathbf{q}|$, $Y_{lm}$ are the spherical harmonics,
and ${\cal I}_{n l}$ is a specific electronic radial integral defined by
\begin{equation}
{\cal I}_{nl}(  q ) =
{\int}_0^{\infty} dr \ r^2 R_{nl}(r) j _{l}(q r) R_{10}(r) -\delta_{n1}\delta_{l0}\delta_{m0}  ,
\label{Inl1}
\end{equation}
where $R_{nl}(r)$ denotes the hydrogenic radial functions  and
$j _{l}(q r) $ are the spherical Bessel functions.
The analytic expression for the specific radial integral on the right-hand side of
  ${\cal I}_{n l}$, Eq. (\ref{Inl1}), is given by Eq. (\ref{Inl}) in    Appendix \ref{A1}.

After  performing the angular integration in Eq. (\ref{def1})
by using the partial-wave expansion of the exponential
term in the generalized form factor
and the definition of ${\mathbf{w}}_{nlm}$ \cite{vf1},
 we obtain for the first term on the right-hand side of the
first-order atomic transition amplitude Eq. (\ref{def1})
\begin{equation}
\langle\psi _{nlm}|\frac{\exp{(i  \mathbf{q} \cdot \mathbf{r} )}}{2 \pi^2 q^2}
| \boldsymbol{\varepsilon}\cdot {\mathbf{w}}_{100}( \Omega_1^+ ) \rangle
=\frac{ i^l  }{{\pi}^{3/2}q^2 }
\left[ \sqrt{\frac{l}{2l+1}}     \; {\cal T}_{n l m}^{l-1,a} (\Omega_1^+,q)
     + \sqrt{\frac{l+1}{2l+1}}   \; {\cal T}_{n l m}^{l+1,a} (\Omega_1^+,q)
\right],
\label{pa}
\end{equation}
\noindent
and for the second term on the right-hand side  of Eq. (\ref{def1}) ,
\begin{equation}
\langle   \boldsymbol{\varepsilon}^*\cdot {\mathbf{w}}_{nlm} ( \Omega_n^-  )
|\frac{ \exp{(i  \mathbf{q} \cdot \mathbf{r} )}}{2 \pi^2 q^2}  |\psi_{100}\rangle
=-\frac{ i^l }{{\pi}^{3/2}q^2 }
\left[ \sqrt{\frac{l}{2l+1}}     \; {\cal T}_{n l m}^{l-1,b}(\Omega_n^-,q)
     + \sqrt{\frac{l+1}{2l+1}}   \; {\cal T}_{n l m}^{l+1,b}(\Omega_n^-,q)
\right],
\label{pb}
\end{equation}
\noindent
where
\begin{eqnarray}
{\cal T}_{n l m}^{l',a} (\Omega_1^+,q)
&=&  \boldsymbol{\varepsilon}  \cdot \mathbf{V}^*_{l' l  m}(\hat {\mathbf{q}})
\;{\cal J}_{nll',1s}^{a}(\Omega_1^+,q),
\label{ta}\\
{\cal T}_{n l m}^{l',b}(\Omega_n^-,q)
&=&  \boldsymbol{\varepsilon}  \cdot \mathbf{V}^*_{l' l  m}(\hat {\mathbf{q}})
\;{\cal J}_{nll',1s}^{b}(\Omega_n^-,q).
\label{tb}
\end{eqnarray}

\noindent
$\mathbf{V}_{l' l m}$    represent the vector spherical harmonics
\cite{varsha}
and ${\cal J}_{nll',1s}^{a} $ and  ${\cal J}_{nll',1s}^{b} $,
with $l'=l\pm1$ (if $l>0$)  and  $l'=1$ (if $l=0$),
are  specific atomic radial integrals  which are defined
by Eqs. (\ref{Ja1})  and  (\ref{Jb1}) in  Appendix \ref{A2}.

The inelastic atomic transition amplitude for a $N$-photon process, $T_{nlm}^{(1)}$,
 is obtained  from Eq. (\ref{tn1}) by substituting
Eqs. (\ref{pa}) and (\ref{pb}) into Eqs. (\ref{def1}) and (\ref{def2})  as

\begin{eqnarray}
T_{nlm}^{(1)}(N) &=&
-\frac{  i^l e^{-i \phi_q} {\cal E}_0 }{2{\pi}^{3/2}q^2 \omega }
\frac{ J_{N-1}(X_q)}{\sqrt{2l+1}}
\left[ \sqrt{l}     \; {\cal T}_{n l m}^{l-1}(\Omega_{1}^+,\Omega_{n}^-,q)
     + \sqrt{l+1}   \; {\cal T}_{n l m}^{l+1}(\Omega_{1}^+,\Omega_{n}^-,q)
\right] \nonumber \\ &&
-\frac{  i^l e^{i \phi_q} {\cal E}_0 }{2{\pi}^{3/2}q^2 \omega }
\frac{ J_{N+1}(X_q)}{\sqrt{2l+1}}
\left[ \sqrt{l}    \; {\cal T}_{n l m}^{l-1} (\Omega_{1}^-,\Omega_{n}^+,q)
     + \sqrt{l+1}  \; {\cal T}_{n l m}^{l+1} (\Omega_{1}^-,\Omega_{n}^+,q)
\right] ,
\label{tni}
\end{eqnarray}
where we have introduced the following notation
\begin{equation}
{\cal T}_{n l m}^{l'}(\Omega_{1}^{\pm},\Omega_{n}^{\mp},q)
=  \boldsymbol{\varepsilon}  \cdot \mathbf{V}^*_{l' l  m}(\hat{\mathbf{q}})
{\cal J}_{nll'}(\Omega_{1}^{\pm},\Omega_{n}^{\mp},q).
\label{t1i}
\end{equation}

\noindent
The specific atomic radial integral ${\cal J}_{nll'} $  is defined as
 the difference between the two atomic radial integrals,  Eqs. (\ref{Ja1}) and (\ref{Jb1}),
\begin{equation}
{\cal J}_{nll'}(\Omega_{1}^{\pm},\Omega_{n}^{\mp},q) \equiv
{\cal J}_{nll'}(\pm\omega,q)=
	 {\cal J}_{nll',1s}^a(\Omega_{1}^{\pm},q)
	-{\cal J}_{nll',1s}^b(\Omega_{n}^{\mp},q),
\label{J}
\end{equation}

\noindent
 with $l'=l\pm1$ (if $l>0$)  and  $l'=1$ (if $l=0$).
The $N$-photon atomic transition amplitude Eq. (\ref{tni})
 involves intermediate states with the angular momentum $l'=l \pm 1$,
where $l$ is the angular momentum of the final state.
We point out that the presence of the phase factors $ e^{i \phi_q}$ and $ e^{- i \phi_q}$
 in  Eq. (\ref{tni}) gives a different kind of interference in  Eq. (\ref{tgen})
between the electronic and atomic transitions amplitudes for CP field in comparison to LP field.

Within the framework described in the previous subsections
the DCS for the inelastic scattering process
accompanied by $N$-photon exchange, summed over the magnetic quantum
number, $m$, of the final state is written in the standard form
\begin{equation}
\frac{d{\sigma}_{nl}(N)}{d\Omega} =
 {(2\pi)}^4  \frac{k_f{(N)}}{k_i} \sum_{m=-l}^{l} {| T_{nlm}(N) |}^2
,\label{dcs}
\end{equation}

\noindent
where the inelastic transition amplitude $T_{nlm}$ is given by
Eq. (\ref{tgen}) together with   Eqs. (\ref{tne})  and (\ref{tni}).
The summation over the magnetic quantum number is
performed   in Eq. (\ref{dcs}) by taking into account the summation formulas
 for the vector spherical harmonics $\mathbf{V}_{l' l m}$ \cite{varsha},
which are presented in   Appendix \ref{A3}.
Finally, after some algebra the DCS for laser-assisted inelastic scattering
$e^-(\textbf{k}_i)$+H($1s$)+$N\omega$ $\to $ $e^-(\textbf{k}_f)$+H($nl$)
by a CP field,
 in which  the energy of the projectile electron is modified by ${ E}_{1s}-{ E}_{n}+ N\omega $,
takes a closed analytic expression,

\begin{eqnarray}
\frac{d{\sigma}_{nl}(N)}{d\Omega }
 &\simeq& \frac{k_f(N)}{k_i} \frac{1}{2 q^4}
\left\{
 8 J_{N}^2(X_q) \;{\cal A}_{nl}(q)  +
\left(\frac{{\cal E}_0}{\omega}\right)^2
|\boldsymbol{\varepsilon}\cdot {\hat \mathbf{q}} |^2 {\cal D}_{nl}(\omega,q)
\right. \nonumber\\&&+
\frac{{ 8 \cal E}_0}{\omega}
|\boldsymbol{\varepsilon}\cdot {\hat \mathbf{q}}| \;  J_{N}(X_q)
[  J_{N-1}(X_q)  \; {\cal B}_{nl}(\omega,q)
+  J_{N+1}(X_q)  \; {\cal B}_{nl}(-\omega,q)]
\nonumber\\&& \left.
 +
\left(\frac{{\cal E}_0}{\omega}\right)^2
[J_{N-1}^2(X_q)\;{\cal C} _{nl}(\omega,q)+J_{N+1}^2(X_q)\;{\cal C} _{nl}(-\omega,q)]
\right\}
,\label{SDN}
\end{eqnarray}

\noindent
where the quantities   ${\cal{A}}_{nl}$,  ${\cal{B}}_{nl}$, ${\cal{C}}_{nl}$
and ${\cal{D}}_{nl}$ are defined as
\begin{eqnarray}
 {\cal A}_{nl}(q)&=& (2l+1) \;{\cal I}_{nl}^2(q),
\label{A}\\
 {\cal B}_{nl}(\omega,q)&=& {\cal I}_{nl}(q)
 \left[ (l+1) {\cal J}_{nll+1}(\omega,q)- l {\cal J}_{nll-1}(\omega,q) \right],
\label{B} \\
 {\cal C}_{nl}(\omega,q)&=&
 \frac{l(l+1)}{2l+1}\left[ {\cal J}_{nll+1}(\omega,q) + {\cal J}_{nll-1}(\omega,q)\right] ^2
 ,\label{C}\\
 {\cal D}_{nl}(\omega,q)&=&
\frac{1}{2l+1} \left\{
{l(l-1)}[{\cal J}_{nll-1}(\omega,q) J_{N-1}(X_q)+{\cal J}_{nll-1}(-\omega,q) J_{N+1}(X_q)]^2
\right. \nonumber\\ &&
+ (l+1)(l+2)[{\cal J}_{nll+1}(\omega,q) J_{N-1}(X_q)+{\cal J}_{nll+1}(-\omega,q) J_{N+1}(X_q)]^2
 \nonumber\\  &&
-{6l(l+1)}[{\cal J}_{nll+1}(\omega,q) J_{N-1}(X_q)+{\cal J}_{nll+1}(-\omega,q) J_{N+1}(X_q)]
\nonumber\\  &&\times\left.
[{\cal J}_{nll-1}(\omega,q) J_{N-1}(X_q)+{\cal J}_{nll-1}(-\omega,q) J_{N+1}(X_q)]
\right\}.
\label{D}
\end{eqnarray}

\noindent

In the particular case of $1s \to ns$ excitation, the DCS  calculated from
Eq. (\ref{SDN}) for $l=0$,  takes a simple analytical expression
\begin{eqnarray}
\frac{d{\sigma }_{n0}(N)}{d\Omega }
 & =& \frac{k_f(N)}{k_i} \frac{1}{q^4}
\left\{
  2 J_{N}(X_q) \;{\cal I}_{n0}(q) \right.
 \nonumber\\ && \left. +
\frac{{\cal E}_0}{\omega} |\boldsymbol{\varepsilon}\cdot {\hat \mathbf{q}}|
 \; [  J_{N-1}(X_q)  \; {\cal J}_{n01}(\omega)
+  J_{N+1}(X_q)  \; {\cal J}_{n01}(-\omega)]
\right\}^2.
\label{SDN1s-n0}
\end{eqnarray}

\noindent
In the domain of negligible laser-atom interaction ($T_{nlm}^{(1)} \simeq 0$),
i.e., at  low-photon energies that are far away from any resonance and weak laser fields,
only the projectile electron-laser interaction has to be considered.
If we neglect the atomic radial integrals in Eqs. (\ref{B})-(\ref{D}),
we obtain a simple formula for the DCS,  Eq. (\ref{SDN}), in the soft-photon limit,
\begin{equation}
\frac{d{\sigma }_{nl}(N)}{d\Omega}=
 \frac{k_f(N)}{k_i}
J_N^2 (  X_q)
\frac{4(2l+1)} {q^4}
\;{\cal I}_{nl}^2(q),
\label{tBFN}
\end{equation}

\noindent
a  result which is equivalent to  the  Cavaliere and Leone \cite{cavaliere}
and Beigman and  Chichkov  formulas  \cite{beigman} derived
for  laser-assisted excitation of hydrogen atoms by fast electrons and LP fields
in which the atomic dressing is neglected.
In particular, the DCS in the soft-photon limit for the ${1s \to 2s}$ excitation
  is calculated using Eq. (\ref{tBFN}) as
\begin{equation}
\frac{d{\sigma }_{20}(N)}{d\Omega}=
 \frac{k_f(N)}{k_i}
J_N^2 ( X_q) \frac{128 }{(9/4+q^2)^6 },
\label{tBFN1s-2s}
\end{equation}
\noindent
and we recover the DCS formula given by Cavaliere and Leone \cite{cavaliere},
\noindent
while for the ${1s \to 2p}$ excitation the DCS  in the soft-photon limit reads
\begin{equation}
\frac{d{\sigma }_{21}(N)}{d\Omega}=
 \frac{k_f(N)}{k_i}
J_N^2 (  X_q )
\frac{288 }{ (9/4+q^2)^6 }\frac{1}{q^2}.
\label{tBFN1s-2p}
\end{equation}
\noindent
The quantities on the right-hand side of  Bessel function
in Eqs.  (\ref{tBFN1s-2s}) and (\ref{tBFN1s-2p}) are related to DCS
for inelastic scattering process in the absence of the laser field \cite{bransden2}.
Therefore the asymptotic forms of DCSs in the soft-photon limit are written as
a product of  the collision kinematics and laser field factors, and field-free DCSs.
The ratio of the  DCSs for  $1s \to 2s$ and $1s \to 2p$ excitations in the soft-photon limit
(when the dressing effect can be negligible),
  given by Eqs. (\ref{tBFN1s-2s}) and (\ref{tBFN1s-2p}), is approximately equal to ${4q^{2}/9}$,
that gives  a larger scattering signal for $1s \to 2p$ in comparison
to the $1s \to 2s$ excitation at small scattering angles.

\subsection{Differential cross section for one-photon exchange ($N= \pm 1$) in the perturbative limit $X_{q} \ll 1 $}

Next, we concentrate our study on the scattering process in a CP laser field
 in which only one photon is exchanged by the colliding system,  and we
derive the formulas for one-photon absorption ($N=1$) and emission  ($N=-1$).
However, we note that the calculations of the $|N| \ge 2 $ processes
at high laser intensities require that the laser-atom interaction should be treated at least
to  second order in the field \cite{acgabi1,krake}.
In the perturbative regime, whenever the argument of the Bessel functions is small, i.e.,
$X_{q} \ll 1$, which is satisfied at low laser intensities or
at small scattering angles with moderate laser intensities,
 the approximate expression of the ordinary Bessel functions
$ J_N(X_{q})\simeq  (X_{q})^N/(2^N {N!})$
  together with $ J_{-N}(X_{q}) = (-1)^N J_{N}(X_{q})$ can be used.
Thus, substituting the first order of the
Bessel functions $ J_{\pm 1}(\alpha_0| \boldsymbol{\varepsilon}\cdot {\mathbf{q}}|)$ in Eq. (\ref{tne}),
the electronic transition amplitude  reads
\begin{equation}
T_{nlm}^{(0)} (N= \pm 1) \simeq \pm  \frac{\sqrt{I} }{(2\pi)^2 }
\frac{|\boldsymbol{\varepsilon}\cdot {\mathbf{q}}|}{ 2 \omega^2 } f_{el}^{B1}(q),
\label{tn0i}
\end{equation}
where $ f_{el}^{B1}(q)$ is the field-free inelastic transition amplitude
in the first-order Born approximation given by Eq. (\ref{fn0i})
and    $I = {\cal E}_0^2$ denotes the laser intensity.

The atomic transition amplitude $T_{nlm}^{(1)} (N= \pm 1)$
is obtained  from Eq. (\ref{tni}) by substituting
the first order of the Bessel functions,
$ J_{0}(\alpha_0| \boldsymbol{\varepsilon}\cdot {\mathbf{q}}|)$,  as
\begin{equation}
T_{nlm}^{(1)}(N=1)\simeq
-\frac{  i^l e^{-i \phi_q}\sqrt{I} }{2{\pi}^{3/2} q^2 \omega  }
\left[ \sqrt{\frac{l}{2l+1}}     \; {\cal T}_{n l m}^{l-1}(\Omega_{1}^+,\Omega_{n}^-,q)
     + \sqrt{\frac{l+1}{2l+1}}   \; {\cal T}_{n l m}^{l+1}(\Omega_{1}^+,\Omega_{n}^-,q)
\right] ,
\label{tn1i}
\end{equation}
for $1s \to nlm$ excitation accompanied by one-photon absorption ($N=1$), and as

\begin{equation}
T_{nlm}^{(1)}(N= -1)\simeq
-\frac{ i^l e^{i \phi_q} \sqrt{I} }{2{\pi}^{3/2}  q^2 \omega}
\left[ \sqrt{\frac{l}{2l+1}}    \; {\cal T}_{n l m}^{l-1} (\Omega_{1}^-,\Omega_{n}^+,q)
     + \sqrt{\frac{l+1}{2l+1}}  \; {\cal T}_{n l m}^{l+1} (\Omega_{1}^-,\Omega_{n}^+,q)
\right] ,
\label{tn-1i}
\end{equation}
\noindent
for one-photon emission ($N=-1$).
 In Eqs. (\ref{tn1i}) and (\ref{tn-1i}) we  keep  the leading term in laser field  for the calculation of
the atomic transition amplitudes and neglect the terms which are proportional to
$ J_{\pm 2}(\alpha_0| \boldsymbol{\varepsilon}\cdot {\mathbf{q}}|)$.

From Eq. (\ref{SDN}) the DCS for the inelastic scattering process in a CP field
accompanied by one-photon exchange  ($N= \pm 1$)
 takes a compact analytic form after some algebra, with a general structure
\begin{equation}
\frac{d{\sigma }_{nl}(N=\pm 1)}{d\Omega }
 \simeq \frac{k_f}{k_i} \frac{I}{2 q^4 \omega^2 }
\left[
{\cal P} _{nl}(\pm \omega,q) +
| \boldsymbol{\varepsilon}\cdot \hat {\mathbf{q}} |^2 {\cal Q}_{nl}( \pm \omega,q)
\right],
\label{SD}
\end{equation}

\noindent
where the terms  ${\cal{P}}_{nl}$ and ${\cal{Q}}_{nl}$ can be expressed by
\begin{equation}
 {\cal P}_{nl}(\omega,q)=
 \frac{l(l+1)}{2l+1}[ {\cal J}_{nll+1}(\omega,q) + {\cal J}_{nll-1}(\omega,q)]^2,
 \label{P}
\end{equation}

\noindent
and
\begin{eqnarray}
 {\cal Q}_{nl}(\omega,q)&=&
\frac{4q}{\omega}\;{\cal I}_{nl}(q)
 \left[  \frac{q}{2\omega} (2l+1)\;{\cal I}_{nl}
 + (l+1) {\cal J}_{nll+1}(\omega,q) - l {\cal J}_{nll-1}(\omega,q) \right]
 \nonumber \\  &&
+ \frac{l(l-1)}{2l+1}{\cal J}_{nll-1}^2(\omega,q)
+ \frac{(l+1)(l+2)}{2l+1} {\cal J}_{nll+1}^2(\omega,q)
 \nonumber \\  && -\frac{6l(l+1)}{2l+1} {\cal J}_{nll+1}(\omega,q)  {\cal J}_{nll-1}(\omega,q)
.
 \label{Q}
\end{eqnarray}

\noindent
Because the initial atomic state is an $s$ state,
 the definitions of quantities ${\cal{P}}_{nl}$ and ${\cal{Q}}_{nl}$ given
by Eqs. (\ref{P}) and (\ref{Q}) have a similar analytical form compared to the definitions
  reported for LP fields  for $H(1s)$ \cite{ac}, as well
  for $H(2s)$ \cite{gabi2s}.
Obviously the expressions of the specific radial integrals
 ${\cal I}_{nl}$ and ${\cal J}_{nll'}$, included in
${\cal{P}}_{nl}$ and ${\cal{Q}}_{nl}$, depend on the
 initial and final states of the inelastic scattering process.
We notice from Eqs. (\ref{SDN}) and (\ref{SD}) that DCS is very sensitive
 to the orientation of the polarization vector with respect to the momentum
transfer vector, which is included in the  scalar product
$ | \boldsymbol{\varepsilon}\cdot \hat{\mathbf{q}}|$.
The DCS has a maximum value in the scattering geometry in which the polarization vector is
parallel to the momentum transfer $\boldsymbol{\varepsilon} \parallel  \mathbf{q}$,
while it takes a minimum value when  $\boldsymbol{\varepsilon} \perp  \mathbf{q}$.
In fact, at low-laser intensities the difference between the
 inelastic one-photon DCSs  for  CP field, Eq. (\ref{SD}),
and that for LP  field \cite{ac,gabi2s} is given by  the polarization term,
$ | \boldsymbol{\varepsilon}\cdot \hat{\mathbf{q}}|^2$,
  which is $ [(\boldsymbol{e}_i \cdot \hat\mathbf{q})^2+(\boldsymbol{e}_j \cdot \hat\mathbf{q})^2]/2 $
for a CP field and $ (\boldsymbol{e}_i\cdot \hat{\mathbf{q}})^2$ for a LP field.

\noindent

In particular, if the dressing of the target  is neglected in  DCS, Eq. (\ref{SD}), i.e.,
the atomic integrals are neglected  (${\cal J}_{nll \pm 1} \simeq 0$)
 in Eqs. (\ref{P}) and (\ref{Q}),   the following simple analytic result is obtained
for the inelastic scattering process accompanied by one-photon exchange
\begin{equation}
\frac{d{\sigma }_{nl}(N= \pm 1)}{d\Omega} \simeq
I \; \frac{ k_f}{k_i}
\frac{|\boldsymbol{\varepsilon}\cdot \hat{\mathbf{q}} |^2} { q^2 \omega^4 }
 ( 2l+1) \; {\cal I}_{nl} ^2 (q).
\label{tBF}
\end{equation}

\subsubsection{Differential cross section for $s$-subshell excitation
 in the perturbative limit  $X_{q} \ll 1 $}

For inelastic scattering  by a CP field where the final state is an $ns$-subshell,
 which means both initial and final atomic states  have spherical symmetry,
the evaluation of  DCS ($1s \rightarrow  ns$) from
 Eqs. (\ref{SD})-(\ref{Q})  with  $l=0$  leads to  the following  expression:
\begin{equation}
\frac{d{\sigma }_{n0}(N= \pm 1)}{d\Omega}=
I \; \frac{k_f}{k_i}
\frac{| \boldsymbol{\varepsilon}\cdot \hat{\mathbf{q}} |^2} {  q^2 \omega^4  }
\left[ {\cal I}_{n0}(q)  \pm  \frac{ \omega}{q }{\cal J}_{n01}(\pm\omega,q) \right]^2
.\label{t1s-ns}
\end{equation}

\noindent
It is worthwhile to mention that  the DCS for  the $n=1$ level given by Eq. (\ref{t1s-ns})
is in excellent agreement with the analytical formulas  derived  for laser-assisted
elastic  electron-hydrogen scattering in the first order laser-atom interaction,
  by  Dubois and coworkers  \cite{dubois1,dubois2} and Cionga  and coworkers  \cite{acgabi2}.
For the $1s \to ns$ excitation process accompanied by one-photon exchange we can easily calculate
the ratio of  DCSs by LP and CP fields from the analytical form of Eq. (\ref{t1s-ns}), as
$| \boldsymbol{\varepsilon_{LP}}\cdot {\mathbf{q}} |^2/
 | \boldsymbol{\varepsilon_{CP}}\cdot {\mathbf{q}} |^2$,
 where  $\boldsymbol{\varepsilon_{LP}}$ and  $\boldsymbol{\varepsilon_{CP}}$
are the corresponding linear and circular polarization vectors.

\noindent
We also present a particular case for  the ${1s \to 2s}$ excitation where the one-photon DCS formula
given by Eq. (\ref{t1s-ns}) can take quite a simple and useful asymptotic expression.
Based on the analytic formula of electronic radial integral ${\cal I}_{20}$, Eq. (\ref{I20}),
 and the low-photon energy limit of atomic radial integral ${\cal J}_{201} $, Eq. (\ref{J201}),
the DCS  for the ${1s-2s}$ excitation  at high projectile electron energies ($E_{k_i}> 200$ eV)
and low-photon energies  ($\omega <0.5$ eV), is given by a simple asymptotic formula
\begin{equation}
\frac{d{\sigma }_{20}(N=\pm 1)}{d\Omega} \simeq
 I \; \frac{k_f}{k_i}
 \frac{| \boldsymbol{\varepsilon}\cdot \hat\mathbf{q} |^2}{ \omega^4 }
\frac{32 }{(9/4+q^2)^6 }\left(q^2 \mp 4.5 \omega +18\omega^2 \right)^2.
\label{t1s-2s}
\end{equation}

\noindent
The first term in the large bracket of Eq. (\ref{t1s-2s}) is connected to the projectile electron
 contribution, while the second and third terms are related to the atomic dressing effect.
At large scattering angles,  where the atomic dressing is negligibly small,
the above equation also gives  accurate results even at photon energies (non-resonant)
in the optical domain.

\subsubsection{Differential cross section for $p$-subshell excitation
in the perturbative limit  $X_{q} \ll 1 $}

For inelastic scattering  by a CP field where the final state is an $np$-subshell,
 which implies that only the initial  atomic state  has a spherical symmetry,
we obtain from  Eqs. (\ref{SD})-(\ref{Q})  evaluated for $l=1$
the following formula for DCS ($1s \rightarrow  np$):
\begin{eqnarray}
\frac{d{\sigma }_{n1}(N= \pm 1)}{d\Omega} &=&
 \frac{k_f}{k_i}
\frac{I}{2 q^4  \omega^2   }
\left\{
\frac{2}{3} ({\cal J}_{n12} +{\cal J}_{n10})^2
 \right.  \nonumber \\ && + \left.
| \boldsymbol{\varepsilon}\cdot \hat{\mathbf{q}} |^2
\left[
\frac{4q}{\omega} {\cal I}_{n1}
\left(\frac{3q}{2\omega}{\cal I}_{n1} \pm 2 {\cal J}_{n12} \mp {\cal J}_{n10} \right)
+
2 {\cal J}_{n12}^2 - 4 {\cal J}_{n12}{\cal J}_{n10}
\right]
\right\} ,
\label{t1s-np}
\end{eqnarray}

\noindent
where for notational simplicity we drop off the arguments, $q$ and $\omega$,
 of the electronic and atomic radial integrals.
Based on the analytic formula of the electronic radial integral ${\cal I}_{21}$, Eq. (\ref{I21}),
 and the low-photon energy limit of atomic radial integrals ${\cal J}_{210} $,
Eq. (\ref{J210}), and ${\cal J}_{212} $, Eq. (\ref{J212}),
the DCS for the ${1s \to 2p}$ excitation calculated  from Eq. (\ref{t1s-np})
at  high  projectile electron energies ($E_{k_i}> 200$ eV) and
low-photon energies  ($\omega <0.5$ eV), is given by a rather simple asymptotic formula
\begin{equation}
\frac{d{\sigma }_{21}(N=\pm 1)}{d\Omega} \simeq
 \frac{k_f}{k_i}  \frac{I}{\omega^4} \frac{72 }{(9/4+q^2)^6 }
\left[  | \boldsymbol{\varepsilon}\cdot \hat\mathbf{q} |^2 (1 \pm 4\omega) + 4 \omega^2
\right].
\label{t1s-2p}
\end{equation}

\noindent
Similar to the $1s \to 2s$ excitation, the terms which are proportional to the photon energy
 $\omega$ and $\omega^2$  in the square bracket of Eq. (\ref{t1s-2p})
 are related to the atomic dressing effect.

\section{Numerical examples and discussion}
\label{III}

\begin{figure}
\centering
\includegraphics[width=2.25in,angle=0]{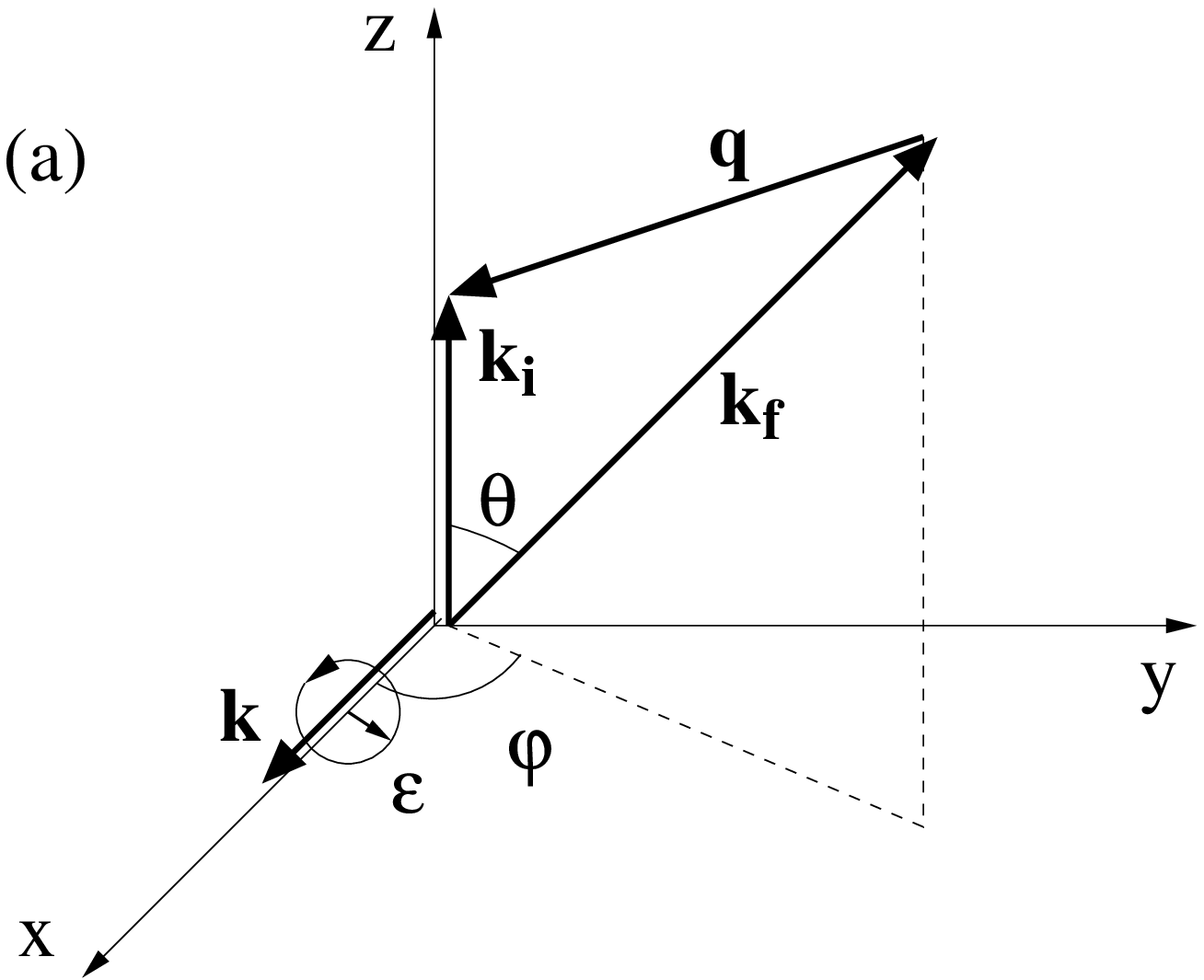}\\
\includegraphics[width=2.25in,angle=0]{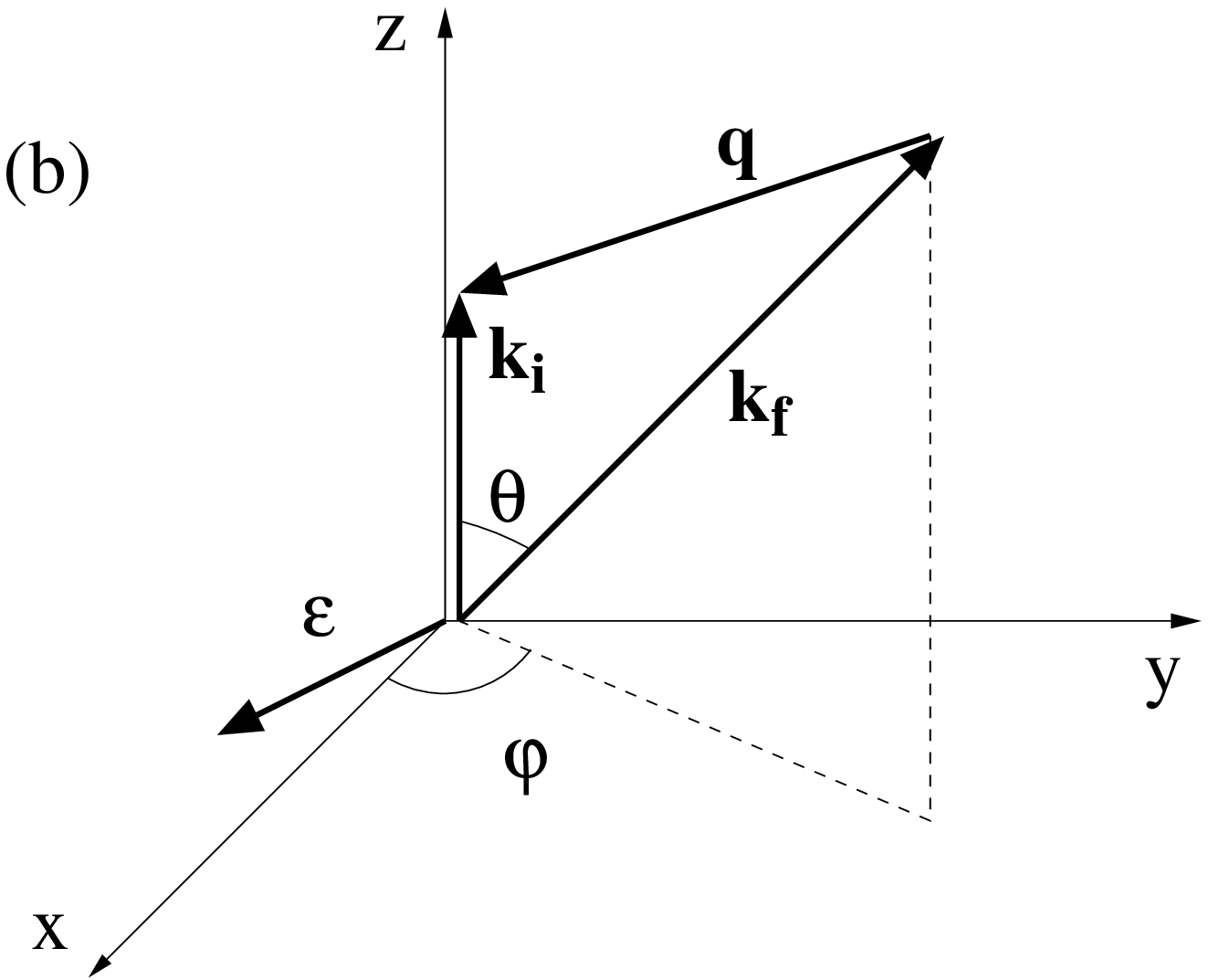}\\
\includegraphics[width=2.25in,angle=0]{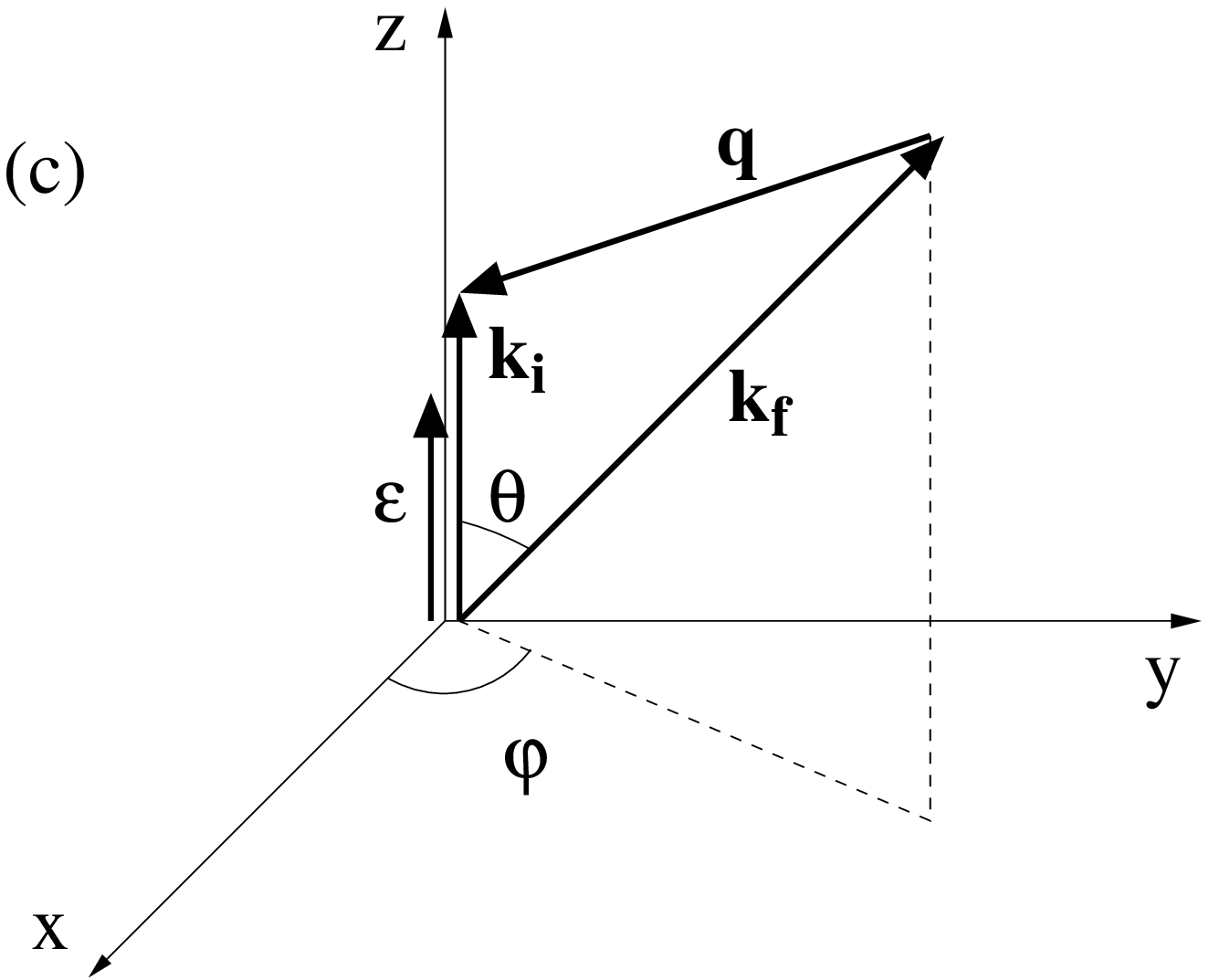}\\
\caption{
The specific scattering geometries  assumed for the numerical
 calculations of laser-assisted $e^-$-H($1s$)  scattering.
$\mathbf{k}_{i}$ and $\mathbf{k}_{f}$ are the initial and final momentum
vectors of the projectile electron, $\theta$ is the angle between them, and
$\mathbf{q} $ is the momentum transfer vector.
$\mathbf{k}$ is the wave vector of the photon,
$\boldsymbol{\varepsilon}$  represents the polarization vector of the
 laser field,  and $\varphi$ is the azimuthal angle.
We consider the scattering geometries with $\mathbf{k}_{i} \parallel Oz$ and
the field polarizations denoted as
(a) \textbf{CP}$_{x}$  where the laser field is circularly polarized, $\bm{\varepsilon_{{CP}_x}}$
lies in the $yz$ plane, and the laser field propagates along the $x$-axis,
(b) \textbf{LP}$_q$   where the laser field is linearly polarized $\bm{\varepsilon_{{LP}_q}} \parallel \mathbf{q}$, and
(c) \textbf{LP}$_z$ where the laser field is linearly polarized $\bm{\varepsilon_{{LP}_z}} \parallel Oz$.
}
\label{fig1}
\end{figure}

We consider the scattering geometries depicted in Figs. \ref{fig1}(a)-\ref{fig1}(c)
 where the momentum vector of the incident electron  $\mathbf{k}_{i}$ is parallel
to the $z$-axis.
 The final momentum of the projectile electron is calculated as
$k_f  (N)= \sqrt{ k_i^2  + 2(E_{1s}- E_{n})  + 2N \omega  }$ and
the amplitude of the momentum transfer vector is given by $q=\sqrt{ k_i^2+ k_f^2 -2 k_i k_f \cos \theta}$,
 where $\theta$ is the the scattering angle between the initial and
final momentum vectors of the projectile electron, ${\bf k}_i$ and ${\bf k}_f$.
We recall that the momentum transfer $q$ varies between a minimum of
 $ |k_i-k_f|$ at forward scattering angles $\theta =0^{\circ}$ and
 a maximum value of $ |k_i+k_f|$ at backward scattering angles $\theta =180^{\circ}$.
We focus our discussion on three particular field polarizations denoted in  Fig. \ref{fig1} as
(a) \textbf{CP}$_{x}$, in which the laser beam is circularly polarized in
the $(y,z)$-scattering plane  and the laser beam propagates in the $x$-axis direction,
$ \bm{\varepsilon_{{CP}_x}}  =(\textbf{e}_y+i \textbf{e}_z)/\sqrt{2}$,
(b)  \textbf{LP}$_q$, in which the laser beam is linearly polarized and the polarization vector is parallel
 the momentum transfer vector, $\bm{\varepsilon_{{LP}_q}} \parallel  \mathbf{q}$, and
(c)  \textbf{LP}$_z$, in which the laser beam is linearly polarized and the polarization vector
 is parallel to the $z$-axis, $ \bm{\varepsilon_{{LP}_z}} =\textbf{e}_z$.
The scalar product, $ \bm{\varepsilon} \cdot \mathbf{q}$, in the argument
of the Bessel function $X_q$ can be simply expressed as
\begin{enumerate}[label=(\alph*)]
\item $ \bm{\varepsilon_{{CP}_x}}\cdot \mathbf{q} =[ i(k_i-k_f \cos \theta) - k_f \sin \theta \sin \varphi]/\sqrt{2}$,
for the \textbf{CP}$_{x}$ polarization,
where $\varphi$ is the azimuthal angle,
\item  $ \bm{\varepsilon_{{LP}_q}} \cdot \mathbf{q}  = q$, for the \textbf{LP}$_q$ polarization,
\item
$ \bm{\varepsilon_{{LP}_z}}\cdot \mathbf{q}  =  k_i-k_f \cos \theta$,
for  the \textbf{LP}$_z$ polarization,
\end{enumerate}

\noindent
and the following ratios hold for the above LP and CP laser fields
\begin{itemize}
 \item
$| \boldsymbol{\varepsilon_{{LP}_q}}\cdot {\mathbf{q}} |^2/
 | \boldsymbol{\varepsilon_{{CP}_x}}\cdot {\mathbf{q}} |^2= 2q/
 [(k_i-k_f \cos \theta)^2 + k_f^2 \sin^2\theta \sin^2\varphi]$,
 \item
$| \boldsymbol{\varepsilon_{{LP}_z}}\cdot {\mathbf{q}} |^2/
 | \boldsymbol{\varepsilon_{{CP}_x}}\cdot {\mathbf{q}} |^2 = 2/
[1 + k_f^2 \sin^2\theta \sin^2\varphi/(k_i-k_f \cos \theta)^2 ]$.
\end{itemize}

First, we checked that the numerical results for one-photon inelastic processes
based on Eq. (\ref{SDN})  are in good agreement with the numerical
 results by Francken and coworkers \cite{Francken},
 calculated for the excitation of the $n=2$ and $3$ levels in hydrogen
in the scattering geometry \textbf{LP}$_q$.
Second, we checked that the numerical results for one-photon inelastic processes
 based on Eq. (\ref{SD}) are in excellent agreement with the previous results
 by Cionga  and Florescu \cite{ac}, for the  excitation  of the $n=4$
 level in hydrogen in the scattering geometries \textbf{LP}$_q$ and \textbf{LP}$_z$.
Finally, we verified the one-photon DCS for the particular case of \textit{elastic} scattering
of fast electrons  by hydrogen atoms in their ground state in the presence of a
CP laser field of  moderate power,  which turn out to be in very good agreement
 with the earlier data \cite{acgabi2}.
We recall that our analytic results are also valid for the case of
 elastic scattering corresponding to the energy conservation
$k_f^2   = k_i^2 + 2N \omega $, a process in which the  hydrogen atom
remains in its ground state and the kinetic energy of the projectile electron
changes by an integer multiple of the photon energy.

\begin{figure}
\centering
\includegraphics[width=3.5in,angle=0]{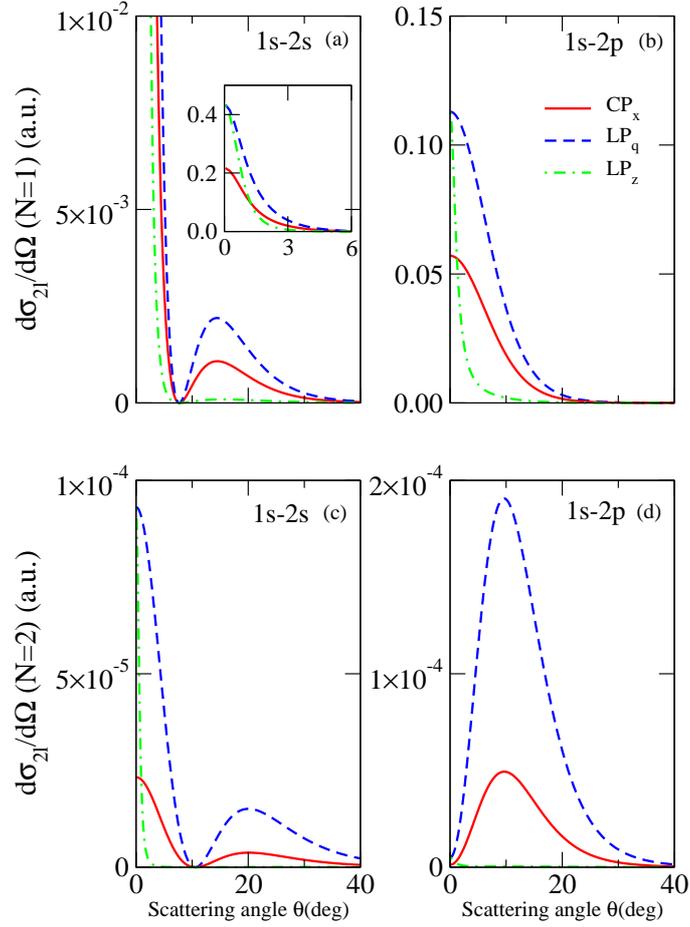}
\caption{(Color online)
Differential cross sections for the laser-assisted
\textit{inelastic} scattering process,  $e^-$+H($1s$) +$N\omega$ $\to$ $e^-$+H($2l$), given by Eq. (\ref{SDN}),
 for $N=1$ in  (a) and (b)  and $N=2$ in  (c) and (d)  as a function of the scattering angle $\theta$,
with excitation of the $2s$ state  in  (a) and (c),  and  $2p$ state  in (b) and (d).
The full lines represent   the scattering geometry \textbf{CP}$_{x}$ with $\varphi=90^{\circ}$,
dashed lines represent the  scattering geometry \textbf{LP$_q$},  and dot-dashed lines represent  the scattering geometry \textbf{LP$_{z}$}.
The initial projectile electron energy is $ E_{k_i}=200$ eV,
the laser field strength is ${\cal E}_0=10^7$ V/cm,
and the photon energy is $ 2 $ eV.
}
\label{fig2}
\end{figure}

Next, we apply the analytic formulas derived in the above sections to calculate the DCSs
  for  inelastic  electron scattering by a hydrogen atom
 in its ground state in the presence of a CP laser field, and
we focus our numerical examples on one- and two-photon absorption.
We choose low-photon energies and high energies of the projectile electron, such that
neither the photon nor the projectile electron can separately excite an upper atomic state.
To start with a simple case we present  the angular distributions in Fig. \ref{fig2}
for an initial scattering energy $ E_{k_i}=200$ eV, a photon energy corresponding
to the He:Ne laser $\omega =2$ eV, a laser field strength ${\cal E}_0=10^7$ V/cm,
 and an azimuthal angle $\varphi=90^\circ$.
These laser field parameters correspond to a value of $\alpha_0 \simeq 0.36$ a.u. and
 $X_q \simeq  0.36 |\boldsymbol{\varepsilon}\cdot {\mathbf{q}}|$.
 The DCSs for one- and two-photon absorption ($N=1$ and $2$),
 calculated from  Eq. (\ref{SDN}), are plotted
 as a function of the scattering angle $\theta$ for
  excitation  of  the $2s$ state in Figs. \ref{fig2}(a) and \ref{fig2}(c),
and  of  the $2p$ state in  Figs. \ref{fig2}(b) and \ref{fig2}(d),
in the scattering geometries \textbf{CP}$_{x}$ (full lines), \textbf{LP}$_q$  (dashed lines),
 and \textbf{LP}$_z$ (dot-dashed lines).
At this laser field strength the two-photon absorption signal is about
three-orders of magnitude smaller than the one-photon absorption signal.
 The DCSs  ($1s \to 2s$) for one-photon absorption at small scattering angles
 [$\theta < 6^{\circ}$ in the inset of Fig. \ref{fig2}(a)],
  show a similar   behavior to  the DCSs for the elastic scattering process
 ($1s \to 1s$), as it is shown in Fig. 1(b$_2$) of Ref. \cite{acgabi2}.
Very recently there are  new experimental detections  for Xe
of the peak profile at forward scattering angles
and a few attempts to explain it based on the Zon's model \cite{Kanya2}.
For the $1s \to 2s$ excitation  the large values of the scattering signal
at $\theta \simeq 0^\circ$ can be simple understood
 from  the analytic expression  of DCS at low-field intensities, Eq. (\ref{t1s-2s}),
as arising from the atomic dressing contribution which is  proportional to  the photon energy,
while the projectile contribution   is proportional to $q^2$
 and can be  negligible at forward scattering angles.

\begin{figure}
\centering
\includegraphics[width=3.5in,angle=0]{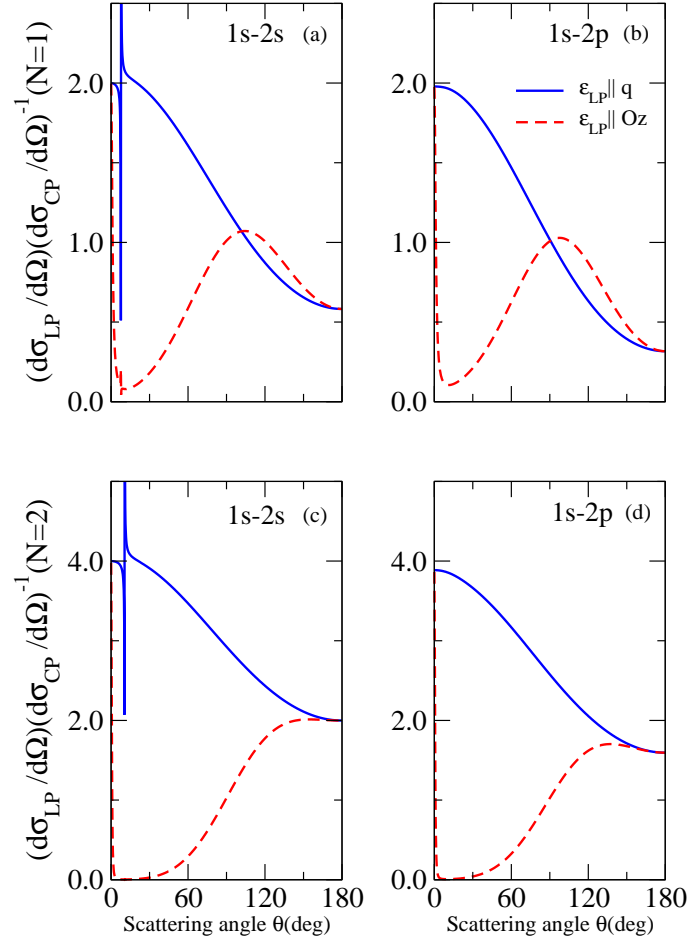}
\caption{
Ratios of the DCSs by the LP and CP fields  for $N=1$ in  (a) and (b)  and $N=2$ in  (c) and (d)
 as a function of the scattering angle $\theta$,
with excitation of the $2s$   and  $2p$ states.
The  full lines represent  the ratio  by the  \textbf{LP}$_{q}$ and  \textbf{CP}$_{x}$ fields,
while the dashed line represent   the ratio  by the  \textbf{LP}$_{z}$ by \textbf{CP}$_{x}$ fields,
 with the  azimuthal angle $\varphi=90^{\circ}$.
The rest of the parameters are the same as in  Fig. \ref{fig2}.
}
\label{fig3}
\end{figure}

Depending of the  laser parameters, projectile energies, and scattering geometry
the CP scattering signal can be larger or smaller than the LP  signal.
In order to better understand the differences between the scattering signals
by the CP and LP fields plotted in Fig. \ref{fig2},  we compare in Fig. \ref{fig3}
the ratios of the DCSs with excitation of the $2s$  and  $2p$ states
 by the LP and CP fields  for $N=1$ in  (a) and (b),  and $N=2$ in  (c) and (d)
 as a function of the scattering angle $\theta$.
 In particular,  in the domain of non-negligible scattering signal  ($\theta <60^{\circ}$)
the DCSs for one- and two-photon absorption  by the \textbf{CP}$_x$ field
are larger than those by the \textbf{LP}$_z$ field and smaller than those by the \textbf{LP}$_{q}$ field.
The ratios of DCSs by the LP and CP fields,
$\frac{d{\sigma }_{2l}(\bm{LP}_{q})}{d\Omega} /
\frac{d{\sigma }_{2l}(\bm{CP}_{x})}{d\Omega}$
and
$\frac{d{\sigma }_{2l}(\bm{LP}_{z})}{d\Omega} /
\frac{d{\sigma }_{2l}(\bm{CP}_{x})}{d\Omega}$,
as a function of the scattering angle
have nearly opposite behavior in Fig. \ref{fig3},
around the value of $1$ for one-photon absorption and $2$ for two-photon absorption.
For $1s \to 2s$  excitation at forward scattering angles
 the \textbf{CP}$_{x}$ signal for one-photon absorption is one-half of the \textbf{LP}$_q$ signal,
as it is displayed in Fig. \ref{fig3}(a) and  in the inset windows of Fig. \ref{fig2}(a).

\noindent
In the perturbative regime ($ X_q \ll 1 $) it can be easily shown from Eq. (\ref{t1s-ns})  that
 the ratio of the one-photon DCSs by the \textbf{LP}$_{q}$ and  \textbf{CP}$_x$ fields
 is equal to $2$ for the $1s \to 2s$ excitation, because for high projectile energies and low-photon energies,
 such that $k_f \simeq k_i$ and $q \simeq 2 k_i \sin (\theta/2) $,  we obtain after simple algebra
$\boldsymbol{\varepsilon_{{CP}_x}}\cdot {\mathbf{q}} \simeq \sqrt 2 k_i \sin (\theta/2) $
at $\varphi=90^\circ$,
that results in a ratio of $| \boldsymbol{\varepsilon_{{LP}_q}}\cdot {\mathbf{q}} |^2/
 | \boldsymbol{\varepsilon_{{CP}_x}}\cdot {\mathbf{q}}|^2\simeq 2$.
At forward scattering angles the \textbf{CP}$_{x}$
signal is also one-half of the \textbf{LP}$_z$ signal, as it is shown in  Fig. \ref{fig3}(a)
 and in the inset window of Fig. \ref{fig2}(a),
because whenever  $\theta \simeq 0^\circ$   the following  ratio holds
$| \boldsymbol{\varepsilon_{{LP}_z}}\cdot {\mathbf{q}} |^2/
 | \boldsymbol{\varepsilon_{{CP}_x}}\cdot {\mathbf{q}}|^2  \simeq 2$.
The sharp peaks at the scattering angles $7.9^{\circ}$ and $10.7^{\circ}$
in Figs. \ref{fig3}(a) and \ref{fig3}(c) occur due to the dynamic minimum of
 $1s \to 2s$  excitation DCSs.
Similarly, for the $1s \to 2p$ excitation, Figs. \ref{fig2}(b) and  \ref{fig3}(b),
it can be shown from Eq. (\ref{t1s-np}) derived
 in the perturbative regime with $ X_q \ll 1 $, that at high projectile energy and
low-photon energy  the ratio of the one-photon DCSs by the \textbf{CP}$_{x}$
(with the azimuthal angle $\varphi=90^\circ$)  and  \textbf{LP}$_q$  fields
 is  close to $2$.
In the case of two-photon absorption, Figs. \ref{fig3}(c)-\ref{fig3}(d),
at very small scattering angles $\theta < 5^{\circ}$,
 the \textbf{CP}$_{x}$ signal is one-quarter of the \textbf{LP}$_q$ signal.
For a $N$-photon process in the perturbative regime ($ X_q \ll 1 $),
we obtain  from  Eq. (\ref{SDN1s-n0}) a simple ratio of DCSs in case of $1s \to ns$ excitation
\begin{equation}
\frac{d{\sigma }_{n0}(\bm{LP}_{q})}{d\Omega} \bigg/
\frac{d{\sigma }_{n0}(\bm{CP}_{x})}{d\Omega}
\simeq 2^{|N|},
\label{ratio}
\end{equation}
 that is independent of the scattering angle and
 is similar to the ratio of DCSs  by the LP and CP fields
for elastic scattering derived in the perturbative domain \cite{acgabi2}.

\begin{figure}
\centering
\includegraphics[width=3.5in,angle=0]{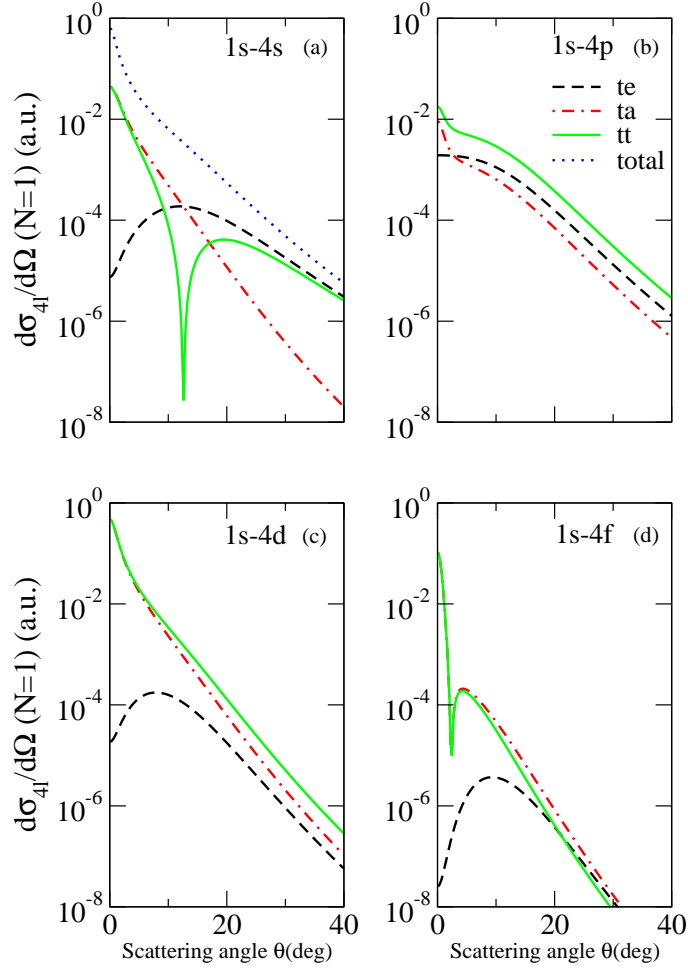}
\caption{(Color online)
Differential cross sections for  laser-assisted \textit{inelastic} scattering process,
$e^-$+H($1s$)+$\omega$ $\to$ $e^-$+H($4l$),
given by Eq. (\ref{SDN}),  with the excitation  of the  $4l$-subshells
 ($l=0,1,2$, and $3$) as a function of the scattering angle $\theta$
for the  (a) $1s\to 4s$,  (b) $1s \to 4p$, (c) $1s\to 4d$, and  (d) $1s\to 4f$ transitions.
The dashed lines represent the  projectile electron contribution,  while the dot-dashed lines
represent the  atomic contribution to DCS.
The scattering geometry is \textbf{CP}$_{x}$ with $\varphi=90^{\circ}$ and
the laser parameters are the same as in Fig. \ref{fig2}.
}
\label{fig4}
\end{figure}

Figures \ref{fig4}(a)-\ref{fig4}(d) show the angular distributions for one-photon absorption ($N= 1$),
 calculated from Eq. (\ref{SDN}) in a logarithmic scale,
 corresponding to the (a) $1s \to 4s$,
(b) $1s \to 4p$,  (c) $1s \to 4d$, and (d) $1s \to 4f$ excitations of hydrogen
 in the scattering geometry \textbf{CP}$_{x}$ with the azimuthal angle  $\varphi = 90^{\circ}$.
The incident projectile energy and laser parameters are the same as in Fig. \ref{fig2}.
The  dashed lines correspond to DCSs  which  neglect
the atomic dressing given by Eq. (\ref{tBFN}) and
the dot-dashed lines represent the atomic contribution to the DCS
calculated as $ {(2\pi)}^4  {k_f}/{k_i} \sum_{m=-l}^{l} {| T_{nlm}^{(1)}(N=1) |}^2 $.
The total inelastic DCS for the excitation of the $n=4$ level is shown
by the dotted line in Fig. \ref{fig4}(a).
The atomic dressing effects are dominant at small scattering angles
for the excitation of the $4l$-subshells ($l=0,1,2$, and $3$),
as well at large scattering angles for the excitation of the $4d$ and $4f$ states only.
Dynamic minimum of DCS in  Fig. \ref{fig4}(a), which is independent of the scattering geometry
 and represents the result of  cancellation of the electronic and
 atomic radial integrals in  Eq. (\ref{SDN}), occurs  for the $1s \to  4s$
 excitation only at the scattering angle $\theta \simeq  12.6^{\circ}$.

\begin{figure}
\centering
\includegraphics[width=5.5in,angle=0]{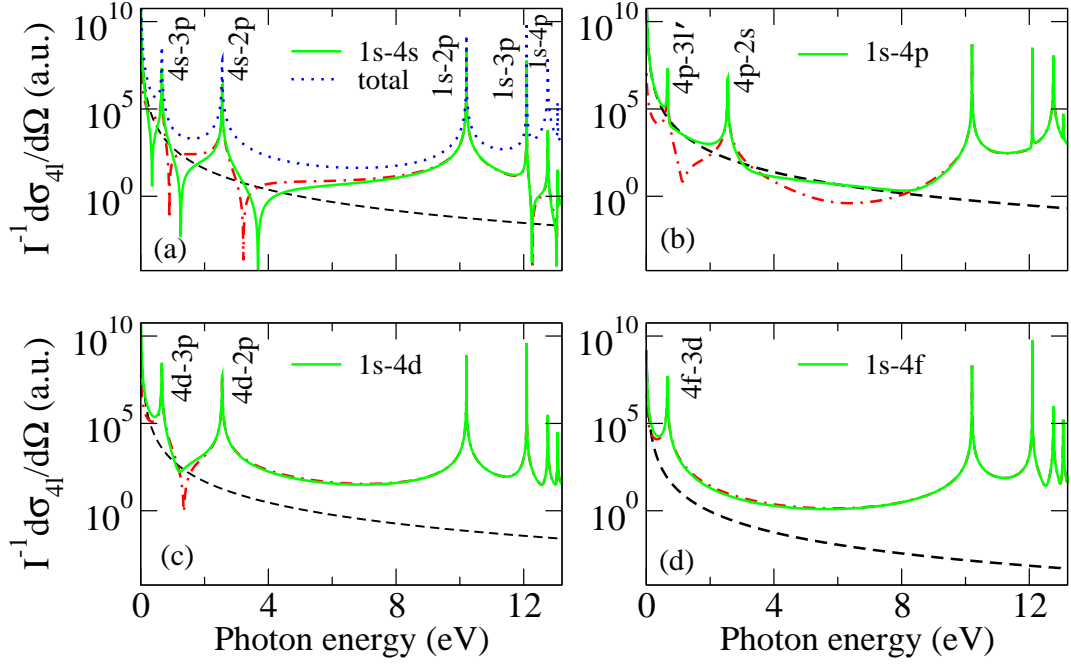}
\caption{(Color online)
Differential cross sections for  laser-assisted \textit{inelastic} scattering process,
$e^-$+H($1s$)+$\omega$ $\to$ $e^-$+H($4l$),
given by Eq. (\ref{SD}),  with the excitation  of the  $4l$-subshells
 ($l=0,1,2$, and $3$) as a function of the laser photon energy
$\omega$, at the projectile electron energy of  $ E_{k_i}=500$ eV and the scattering angle of $5^{\circ}$,
for the  (a) $1s\to 4s$,  (b) $1s \to 4p$, (c) $1s\to 4d$, and  (d) $1s\to 4f$ transitions.
The dashed lines represent the  projectile electron contribution,  while the dot-dashed lines
represent the  atomic contribution to DCS.
The dotted line in Fig. \ref{fig5}(a) represents the total DCS for the excitation of the $n=4$ level.
The scattering geometry is \textbf{CP}$_{x}$ with $\varphi=90^{\circ}$ and the DCSs are normalized by the laser intensity $I$.
}
\label{fig5}
\end{figure}

\begin{figure}
\centering
\includegraphics[width=3.75in,angle=0]{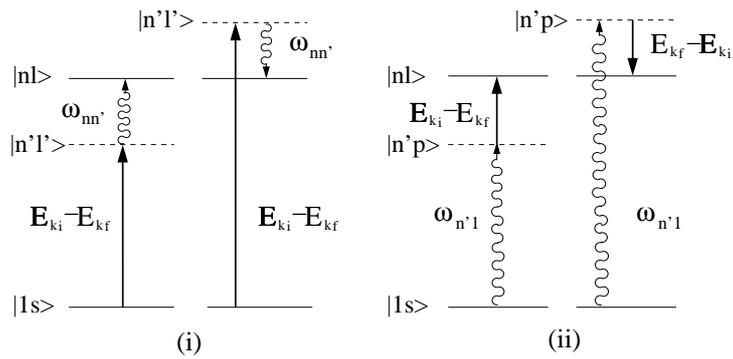}
\caption{(Color online)
Schematic energy level diagrams relevant for the
simultaneously projectile electron-photon excitation of  hydrogen to an $nl$ ($n>2$) state either
(i) by first absorbing some of the projectile electron kinetic energy,
 then absorbing (emitting) a photon from (to) the laser field
or (ii) by first absorbing  a photon from the laser field, then absorbing (transferring)
the energy difference from (to) the projectile electron.
}
\label{fig6}
\end{figure}

At low laser intensities we show in Figs \ref{fig5}(a)-\ref{fig5}(d) the DCSs,
given by Eq. (\ref{SD}),
 with respect to the laser photon energy for one-photon absorption ($N= 1$)
corresponding to the (a) $1s \to 4s$, (b) $1s \to 4p$,  (c) $1s \to 4d$,
 and (d) $1s \to 4f$ excitations of hydrogen  in the scattering geometry \textbf{CP}$_{x}$.
The DCSs are normalized to the laser intensity and calculated
  at a higher incident projectile electron energy  $ E_{k_i}=500$ eV,
a  scattering angle  $\theta = 5^{\circ}$, and an azimuthal angle  $\varphi = 90^{\circ}$.
The  dashed lines correspond to DCSs  which  neglect
the atomic dressing given by Eq. (\ref{tBF}) and
the dot-dashed lines represent the atomic contribution to the DCS.
The total inelastic DCS for the excitation of the $n=4$ level is plotted
by the dotted line in Fig. \ref{fig5}(a).
The DCSs for $1s \to 4l$ ($l=0,1,2$, and $3$) excitations as a function of the photon energy
show a strongly dependence on the atomic structure and exhibits two sets of resonances.
These two sets of resonances generally occur,
as schematically plotted in  Fig. \ref{fig6},  due to the simultaneous
projectile electron-photon excitation of  hydrogen  to the $4l$ state either
(i) by first absorbing some of the projectile electron kinetic energy,
 then absorbing (emitting) a photon from (to) the laser field
or (ii) by first absorbing  a photon from the laser field, then absorbing (transferring)
the energy difference from (to) the projectile electron.
The first set of resonances in Figs. \ref{fig5}(a)-\ref{fig5}(d) occurs
at the photon energy $\omega \simeq 0.66 $ eV, that matches the energy difference
 between  the $4s $ and  $ 3p$ states.
The origin of this resonance resides in the dipole  coupling  of a final $4l$ state
and a lower intermediate $3l'$ state ($l' =l \pm 1$ with $l'<3$).
The second resonance with a similar origin occurs  at $\omega \simeq 2.55 $ eV
 in Figs. \ref{fig5}(a)-\ref{fig5}(c) due to the dipole  coupling
of a final $4l$ state and a lower intermediate $2l'$ state ($l' =l \pm 1$ with $l'<2$).
As it can be noted from Eq. (\ref{pb}) this first set of resonances
 is related to the poles of the atomic transition amplitude due to
the atomic radial integral  $ {\cal J}^{b}_{4ll',1s}(\Omega_{4}^-,q)$, given by Eq. (\ref{Jb1i}),
 at photon energies such that $\omega_{4n'} =  (1/{n'}^{ 2}-1/4^2) /2$, with $ n' =2$ and $3$.
The second set of resonances that appear in Figs. \ref{fig5}(a)-\ref{fig5}(d),
at photon energies  $\omega \simeq 10.2, 12.09, 12.75$ eV,...,
 are associated to the $1s \to n'p$ atomic transitions ($n' =2,3,4,...$).
These resonances occur due to one-photon absorption from
the initial $1s$ state to an intermediate  $n'p$  state, followed by a transition
 to the final $4l$ state  by projectile electron interaction,
as it can be understood from the atomic transition amplitude Eq. (\ref{pa}) and  Fig. \ref{fig6}(ii).
For this second set of resonances the other atomic radial integral $ {\cal J}^{a}_{4ll',1s}(\Omega_{1}^+,q)$,
given by Eq. (\ref{Ja1i}),  presents poles that occur at photon energies that match  atomic resonances
such that $\omega_{n'1} = (1-1/{n'}^{2}) /2$, with $n' \ge 2$.
We remark that, despite the new experimental progresses on laser-assisted electron-atom scattering
   \cite{Harak,Kanya}, such investigations are still  quite difficult to accomplish,
especially for those involving excited states of atoms.
From the experimental point of view we believe that some of these
resonances are particularly interesting \cite{Dunseath2011}, and it
 may be feasible  to detect the inelastic process at photon energies close
 to the $0.66$- or $2.55$-eV resonances because these values do not correspond to any resonance
 of the \textit{elastic} scattering processes of hydrogen from its ground state  \cite{dubois1}.

\section{Summary and conclusions}
\label{IV}
In this paper study the influence of laser polarization in
 electron-impact excitation of hydrogen atoms in a circularly polarized laser field.
Using a semi-perturbative method new analytic formulas in a closed form have been derived for the
 DCSs in laser-assisted inelastic  electron-hydrogen scattering  for $1s \to nl$ excitation,
which are valid for both linear and circular polarization, with
the kinematic part that depends on the scattering geometry and photon polarization vector
clearly separated.
 We obtained a very good agreement with the earlier numerical results for LP fields,
showing the accuracy and efficiency of our theoretical results.
A comparison between the linear and circular polarizations of the laser field
was made for different scattering geometries for the excitation of the $n=2$ levels
 and important differences occur between the  scattered electron
angular distributions depending on the type of polarization:
 \textbf{CP}$_x$,  \textbf{LP}$_q$, and  \textbf{LP}$_z$.
From the extensive calculations we have obtained that, at non-resonant photon energies
in the domain of non-negligible scattering signal (at scattering angles $\theta<60^\circ$),
 the DCSs  for one- and two-photon excitation  of the $2s$ and $2p$ states
by the \textbf{CP}$_x$ field  are generally larger  than  by the \textbf{LP}$_z$ field,
and smaller  than by the \textbf{LP}$_q$ field.
The  detailed numerical data presented for the excitation of the $n=4$ levels by CP fields
indicate that  the atomic dressing effects for inelastic processes
  are important and we found  a significant increase in the DCSs at small scattering angles
for the  $s-s$, $s-d$, and $s-f$ optically forbidden transitions
due to  the simultaneous electron-photon excitation.
We have also elucidated the origin of the peaks in the resonance structure
 of  DCSs as occurring due to  the dipole coupling of
the initial ground state with  $n'p$  states and of
the final excited state with   $n'l'$  ($l'=l\pm 1$) states.
It was found that by changing the laser field polarization the angular distribution
and the photon frequency dependence of the scattering signal can be modified.

\newpage

\begin{appendices}
\section{Analytic expression for the electric radial integral ${\cal I}_{nl} $ }
\label{A1}

We recall that the electronic radial integral, $ {\cal I}_{nl}$,
 needed for the calculation of the inelastic electronic
transition amplitude, $T_{nlm}^{(0)}$,  is defined as
\begin{equation}
{\cal I}_{nl}( q ) =
{\int}_0^{\infty} dr \ r^2 R_{nl}(r) j _{l}(q r) R_{10}(r) -\delta_{n1}\delta_{l0}.
\end{equation}

\noindent
Different methods can be used for the calculation of the electronic radial integral \cite{whelan},
and an analytic expression for the electronic radial integral is obtained,
after performing the integration over the projectile coordinates,
as a finite sum of Gauss hypergeometric function, $_2F_1$:
\begin{eqnarray}
{\cal I}_{nl}(q) &=&
\frac{1 }{q}
\frac{2^{l+2}}{ (2l+1)!}
\left[\frac{(n+l)!}{(n-l-1)!}\right]^{1/2}
 \nonumber \\ && \times
 {\rm Re}  \left\{ \sum_{p=0}^{l}
\frac{i^{p-l-1}}{(2 q)^{p}}
\frac{(l+p)!}{p!(l-p)!}
\frac{(l+1-p)!}{  (1+n- i qn)^{2+l-p} \; n ^p} \right.
\nonumber\\ && \times  \left.
 _2F_1(l+2-p,l-n+1,2l+2,\frac{2 }{1+n- i qn})
\right\} -\delta_{n1}\delta_{l0}.
\label{Inl}
\end{eqnarray}

\noindent
For the particular case of  $1s \to 2s$ excitation, the electronic integral ${\cal I}_{20}$
calculated from Eq. (\ref{Inl}) is given by
\begin{equation}
{\cal I}_{20}(q)= \frac{4\sqrt{2} \; q^2}{(9/4+q^2)^3 }\;,
\label{I20}
\end{equation}
\noindent
\noindent
while for the $1s \to 2p$ excitation, the electronic radial integral ${\cal I}_{21}$   reads
\begin{equation}
{\cal I}_{21}(q)= \frac{2\sqrt{6}\; q }{(9/4+q^2)^3 }\;,
\label{I21}
\end{equation}
where the expressions of  ${\cal I}_{20}$  and  ${\cal I}_{21}$  agree with  the formulas given by Bransden and Joachain \cite{bransden2}.

\section{Analytic expressions for the atomic radial integrals
$ {\cal J}^{a}_{nll',1s}$  and $ {\cal J}^{b}_{nll',1s}$}
\label{A2}

In order to evaluate the  atomic transition amplitude $T_{nlm}^{(1)}$
the atomic radial integral ${\cal J}_{nll'} $, Eq. (\ref{J}),
  is rewritten as  the difference between two atomic radial integrals,
\begin{equation}
 {\cal J}_{nll'}({\pm}\omega,q) \equiv
{\cal J}_{nll'}(\tau_{1}^{\mp},\tau_{n}^{\pm},q)
= 	 {\cal J}_{nll',1s}^a(\tau_{1}^{\mp},q)
	-{\cal J}_{nll',1s}^b(\tau_{n}^{\pm},q) \;,
\label{J1}
\end{equation}
with the selection rule $l'=l \pm 1 $,
where the dependence of the atomic radial integrals on the
energies $\Omega$ is now included in the new parameters
$ \tau^{+}_n= 1/\sqrt{-{2\Omega^{-}_n}}$ and
$ \tau^{-}_n =1/\sqrt{-{2\Omega^{+}_n}}$.
The two  radial integrals on the right-hand side of Eq. (\ref{J1}) are defined as,
\begin{equation}
{\cal J}_{nll',1s}^a( \tau_1, q ) =
{\int}_0^{\infty} dr\ r^2 R_{nl}(r) j _{l'}(qr)
{\cal B}_{101} ( \tau_1 ; r ) \; ,
\label{Ja1}
\end{equation}
and
\begin{equation}
{\cal J}_{nll',1s}^b( \tau_n, q ) =
{\int}_0^{\infty} dr\ r^2 R_{10}(r) j _{l'}(q r)
{\cal B}_{nll'} ( \tau_n ; r ) \;
,\label{Jb1}
\end{equation}
 where the radial functions ${\cal B}_{nll'}( \tau ; r )$  are defined  in Ref. \cite{vf1}.

By using the development of the spherical Bessel function in Eq. (\ref{Ja1}),
 after performing some calculations, we obtained an analytically expression for
the  atomic radial integral ${\cal J}_{nll',1s}^a$  in terms of finite sums of the
 Appell hypergeometric functions \cite{appell}, $F_1$,  of two variables,
\begin{eqnarray}
{\cal J}_{nll',1s}^a(\tau,q)&=&
\frac{2^{\tau +l +4} \; \tau }{q  \; n^{2+l} (1+\tau)^{2+\tau}(1-\tau)^{2}(2-\tau)}
\frac{1}{(2l+1)!}
\left[\frac{(n+l)!}{(n-l-1)!}\right]^{1/2}
\nonumber \\ && \times
{\rm Re} \left\{ \sum_{p=0}^{l'}
\frac{i^{p-l'-1}}{(2 q)^{p}}
\frac{(l'+p)!}{p!(l'-p)!}
\sum_{s=0}^{n-l-1}
\frac{(l+1-n)_s}{(2l+2)_s s!} \left(\frac{2}{n}\right)^s
\right. \nonumber \\ && \times
(2+l-p+s)!
\left(\frac{n\tau } {n+\tau-i q n\tau}\right)^{3+l-p+s}
\nonumber\\ && \times  \left.
 F_1(2-\tau,-1-\tau ,3+l-p+s, 3-\tau,x_1,y_1)
\right\}
\label{Ja1i}.
\end{eqnarray}

\noindent
The term $(p)_s$, with $p$ and $s$ integers, denotes the Pochhammer's symbol
and   the Appell hypergeometric functions $F_1$ depend on the variables
\begin{equation}
x_1 =\frac{1-\tau}{2}  , \;
y_1 =  \frac{n(1-\tau)} {n+\tau-i q n\tau} \;.
\end{equation}
\noindent
Note that ${\cal J}_{nll',1s}^a$, Eq. (\ref{Ja1i}), presents poles with respect to  $\tau$
 which  arise due to the cancellation  of the $n'-\tau $ ($n'=1,2 $)  denominators and
 from the poles of the Appell hypergeometric functions  for $\tau = n'$, where $n'$ is an integer ($n' > 2 $).
The origin of these poles resides in the poles of the Coulomb Green's functions
employed for the calculation of $\textbf{w}_{nlm}$ vectors \cite{vf1}.
The second set of one-photon resonances discussed in Figs. \ref{fig5}(a)-\ref{fig5}(d) is related to the poles
of the  radial integral  ${\cal J}_{nll',1s}^a( \tau_1^{\pm}, q ) $  at
at $\tau_1^- = n'$  with $n'>1$.

We consider now the other atomic radial integral $ {\cal J}_{nll',1s}^b$ defined
 in Eq. (\ref{Jb1}) which is calculated, in the same manner as $ {\cal J}_{nll',1s}^a$,
 in terms of finite sums  of Appell hypergeometric functions, $F_1$, as
\begin{eqnarray}
{\cal J}_{nll',1s}^b(\tau,q)&=&
\frac{ \tau }{q} \;
\frac{2^{2l'+\tau +2} \; n^{\tau-1} }{(2l'+1)!}
\left[\frac{(n+l)!}{(n-l-1)!}\right]^{1/2}
  {\rm Re} \left\{ \sum_{k=-1,1} d_{\; n,l}^{\; l',-k} \right.
\nonumber \\ && \times
 \sum_{p=0}^{l'}
\frac{i^{p-l'-1}}{(2 q)^{p}}
\frac{(l'+p)!}{p!(l'-p)!}
\sum_{\nu=0}^{n-l'-1-k}
\frac{(4)^\nu}{\nu!} (l'+1-p+\nu)!
\nonumber \\ && \times
\sum_{\mu=0}^{n-l'-1-k-\nu}
\left(\frac{-1}{2n}\right)^\mu
\frac{(l'+1+k-n)_{\mu+\nu}}{b \; (2l'+2)_{\nu} \mu! }
 \nonumber \\ && \times
\frac{(n+\tau)^{k+2\mu-n-\tau} }{(n-\tau)^{k+\mu+\nu-n+l'+1} }
\left(\frac{\tau}{1+\tau- iq\tau}\right)^{2+l'-p+\nu}
\nonumber\\ && \times  \left.
 F_1(b,-n-\tau+1+k +\mu,2+l'-p+ \nu, b+1,x_n,y_n)
\right\},\label{Jb1i}
\end{eqnarray}
\noindent
where the  expressions
$d_{\; n,l}^{\; l+1,1} =(n+l+1)(n+l+2) $,
$d_{\; n,l}^{\; l-1,1} = 1$, and
$d_{\; n,l}^{\; l',-1} = - d_{-n,l}^{ \ l',1} $
 are  defined  in  \cite{vf1},
 $b =l'+1-\tau + \mu+\nu $,
 and  the Appell hypergeometric functions $F_1$ depend on other two variables
\begin{equation}
x_n =\frac{n-\tau}{2n},
\;
y_n= \frac{n-\tau}{n(1+\tau-i q \tau)}.
\end{equation}
\noindent
Compared to $ {\cal J}^{a}_{nll',1s}$, the atomic radial integral
$ {\cal J}^{b}_{nll',1s}$ presents poles with respect to  $\tau$  in Eq. (\ref{Jb1i}), which arise due to the
cancellation of the $n -\tau$ and $b$ denominators, as well as from
the poles of the Appell hypergeometric function $F_1$   for $\tau = n'$, where $n'$ is an integer.
The first set of one-photon resonances discussed in Figs. \ref{fig5}(a)-\ref{fig5}(d) is related to the poles
of the  radial integral  ${\cal J}_{nll',1s}^b( \tau_n^{\pm}, q )$ at
$\tau_n^- = n'$ with $n'>n$ (for emission) and $\tau_n^+ = n'$ with $n'<n$ (for absorption).

In addition, for the $1s \to 2s$ excitation the atomic radial integral ${\cal J}_{201}(\omega,q) $
 calculated from Eqs. (\ref{J1}), (\ref{Ja1i}), and  (\ref{Jb1i})
 is approximated in the low-photon energy limit  ($\omega \ll E_{n}-E_{1s}$) as
\begin{equation}
 {\cal J}_{201}(\omega,q) \simeq
-  q\frac{18 \sqrt{2}}{(9/4+q^2)^3}
+q  \; \omega \frac{3\sqrt{2} (243+684q^2+128q^4)}{2(9/4+q^2)^5},
\label{J201}
\end{equation}

\noindent
while for the $1s \to 2p$ excitation we obtain simple formulas
in the low-photon energy limit of ${\cal J}_{210}(\omega,q) $ and  ${\cal J}_{212}(\omega,q) $
\begin{eqnarray}
 {\cal J}_{210}(\omega,q) &\simeq&
-  {q^2}\frac{12 \sqrt{6}}{(9/4+q^2)^3}
-q  \; \omega \frac{\sqrt{6} (1701+3780q^2-336q^4-64q^6)}{16(9/4+q^2)^5},
\label{J210}\\
 {\cal J}_{212}(\omega,q) &\simeq&
q^2   \omega \frac{18\sqrt{6} (21+4q^2)}{(9/4+q^2)^5},
\label{J212}
\end{eqnarray}

\noindent
where the neglected terms are of $\omega^2$ order.
These asymptotic expressions are very useful to provide simple approximate formulas for
$1s \to 2s$ and $1s \to 2p$ excitation DCSs, Eqs. (\ref{t1s-2s}) and (\ref{t1s-2p}), and
to verify the numerical results from Eqs. (\ref{t1s-ns}) and (\ref{t1s-np}), as well.

\section{Some useful summation formulas for the vector spherical harmonics,  $\textbf{V}_{l' l m} $ }
\label{A3}

The well-known summation formulas  of the vector spherical harmonics \cite{varsha}, $\mathbf{V}_{l\pm1 l m}$,
used for the calculation of the DCS, Eq. (\ref{SDN}), and of the quantities $B_{nl}$, Eq. (\ref{B}), $C_{nl}$, Eq. (\ref{C}), and $D_{nl}$, Eq. (\ref{D}), are presented below:

\begin{eqnarray}
\sum_{m=-l}^{l}
[\textbf{a}_1 \cdot \textbf{V}_{l+1 l m}^*(\hat \textbf{q})]
[\textbf{a}_2 \cdot \textbf{V}_{l+1 l m}(\hat \textbf{q})]
&=&
 \frac{1}{8 \pi} \left[ l \textbf{a}_1 \cdot \textbf{a}_2
 + (l+2)(\textbf{a}_1 \cdot \hat \textbf{q})(\textbf{a}_2 \cdot \hat \textbf{q}) \right],
\\
\sum_{m=-l}^{l}
[\textbf{a}_1 \cdot \textbf{V}_{l-1 l m}^*(\hat \textbf{q})]
[\textbf{a}_2 \cdot \textbf{V}_{l-1 l m}(\hat \textbf{q})]
&=&
 \frac{1}{8 \pi} \left[ (l+1)  \textbf{a}_1 \cdot \textbf{a}_2
 + (l-1)(\textbf{a}_1 \cdot \hat \textbf{q})(\textbf{a}_2 \cdot \hat \textbf{q}) \right],
\\
\sum_{m=-l}^{l}
[\textbf{a}_1 \cdot \textbf{V}_{l+1 l m}^*(\hat \textbf{q})]
[\textbf{a}_2 \cdot \textbf{V}_{l-1 l m}(\hat \textbf{q})]
&=&
 \frac{\sqrt{l(l+1)}}{8 \pi} \left[ \textbf{a}_1 \cdot \textbf{a}_2
 -3  (\textbf{a}_1 \cdot \hat \textbf{q})(\textbf{a}_2 \cdot \hat \textbf{q}) \right],
\\
\sum_{m=-l}^{l} Y^*_{lm}(\hat \textbf{q}) \textbf{V}_{l+1 l m}(\hat \textbf{q})&=& -\frac{\sqrt{(l+1)(2l+1)}}{4 \pi} \;\;\hat \textbf{q},
\\
\sum_{m=-l}^{l} Y^*_{lm}(\hat \textbf{q}) \textbf{V}_{l-1 l m}(\hat \textbf{q})&=&\frac{\sqrt{l(2l+1)}}{4 \pi}
\;\hat \textbf{q}.
\end{eqnarray}
\end{appendices}

\section*{Acknowledgments}
The work by G. Buica was supported by research program
Laplace IV contract PN 16 47 02 02 and contract Capacitati FAIR-RO 07-FAIR/2016
from the ANCSI and the Ministry of Education and Research (Romania).

\bibliographystyle{elsarticle-num}

\clearpage
\newpage

\end{document}